\documentstyle[12pt,preprint]{aastex}

\def\arcsecpoint{$''\!.$}

\slugcomment{to appear in {\it The Astrophysical Journal}}

\lefthead{Kraemer, Crenshaw, Hutchings, George, Danks, Kaiser, Nelson, \&
Weistrop}
\righthead{STIS Echelle Observations of the Seyfert Galaxy NGC 4151:
Physical Conditions in the Ultraviolet Absorbers}

\begin{document}

\title{STIS Echelle Observations of the Seyfert Galaxy NGC 4151:
Physical Conditions in the Ultraviolet Absorbers\altaffilmark{1}}

\author{ S. B. Kraemer\altaffilmark{2},
D. M. Crenshaw\altaffilmark{2},
J. B. Hutchings\altaffilmark{3},
I.M. George\altaffilmark{4,5},
A.C. Danks\altaffilmark{6},
T.R. Gull\altaffilmark{7},
M.E. Kaiser\altaffilmark{8},
C.H. Nelson\altaffilmark{9},
\& D. Weistrop\altaffilmark{9}}

\altaffiltext{1}{Based on observations made with the NASA/ESA Hubble Space 
Telescope. STScI is operated by the Association of Universities for Research in 
Astronomy, Inc. under the NASA contract NAS5-26555. }

\altaffiltext{2}{Catholic University of America,
NASA's Goddard Space Flight Center, Code 681,
Greenbelt, MD  20771; stiskraemer@yancey.gsfc.nasa.gov, 
crenshaw@buckeye.gsfc.nasa.gov.}

\altaffiltext{3}{Dominion Astrophysical Observatory,
National Research Council of Canada, 
Victoria, BC V8X 4M6, Canada;
John.Hutchings@hia.nrc.ca.}

\altaffiltext{4}{Laboratory for High Energy Astrophysics,
NASA's Goddard Space Flight Center, Code 662,
Greenbelt, MD  20771; george@oberon.gsfc.nasa.gov}

\altaffiltext{5}{Joint Center for Astrophysics,
University of Maryland, Baltimore County, 1000 Hilltop Circle,
Baltimore, MD 21250}

\altaffiltext{6}{Raytheon Polar Services,
NASA's Goddard Space Flight Center, Code 681,
Greenbelt, MD  20771; danks@stars.gsfc.nasa.gov.}

\altaffiltext{7}{NASA's Goddard Space Flight Center, Code 681,
Greenbelt, MD  20771; gull@sea.gsfc.nasa.gov.} 

\altaffiltext{8}{Department of Physics and Astronomy,
Johns Hopkins University,
3400 North Charles Street,
Baltimore, MD  21218; kaiser@munin.pha.jhu.edu.} 

\altaffiltext{9}{Department of Physics,
University of Nevada, Las Vegas,
4505 Maryland Parkway,
Las Vegas, NV 89154-4002; cnelson@physics.unlv.edu,
weistrop@nevada.edu.}

\begin{abstract}

We have examined the physical conditions in intrinsic UV-absorbing gas
in the Seyfert galaxy NGC 4151, using echelle spectra obtained with the
Space Telescope Imaging Spectrograph (STIS) on the {\it Hubble Space Telescope
(HST)} on 1999 July 19. We confirm the presence of the
kinematic components detected in earlier Goddard High Resolution Spectrograph
(GHRS) observations, all of which appear to be outflowing from the nucleus, 
as well as a new broad absorption feature at a radial
velocity of -1680 km s$^{-1}$. The UV continuum of NGC 4151 was a factor of 
about 4 lower than in observations taken over the previous two years, and 
we argue the changes in the column density of the low ionization 
absorption lines associated with the broad component at -490 km s$^{-1}$ reflect the decrease in the ionizing flux.
Most of the strong absorption lines (e.g., N~V, C~IV, Si~IV, etc.) from this 
component are saturated, but show substantial 
residual flux in their cores, indicating that the absorber does 
not fully
cover the source of emission. Our interpretation is that the unocculted light is due
to scattering by
free electrons from an extended region, which reflects continuum, emission 
lines, and absorption lines. For the first time in such a study, we have been 
able to constrain the densities for this kinematic component and several
others based on the 
strength of absorption lines from metastable states of C~III and Fe~II, and/or
the ratios of ground and fine structure lines of O~I, C~II, and Si~II.
We have generated a set of photoionization models  
which not only successfully match
the ionic column densities for each component during the present low flux 
state, but 
also those seen in previous high flux states with the GHRS and STIS, confirming
that the absorbers are photoionized and respond to the changes in the
continuum flux. 
Based on the model parameters (ionization parameter and density),
we have been 
able to map the relative radial positions of the absorbers. We find that 
the absorbing gas 
decreases in density with distance. Finally, none of the UV absorbers is of sufficiently large column
density or high enough ionization state to account for the observed
X-ray absorption, while the scatterer is too highly ionized. Hence, the X-ray
absorption must arise in a separate component of circumnuclear gas.

\end{abstract}


\section{Introduction}

NGC~4151 (cz $=$ 995 km s$^{-1}$; determined from
21 cm observations, de Vaucouleurs et al. [1991]) is the first Seyfert galaxy known to show
intrinsic absorption that could be attributed to the active nucleus. 
Ultraviolet observations of NGC~4151 by the {\it IUE} (Boksenberg et al. 
1978) and subsequent far-ultraviolet observations by the {\it Hopkins 
Ultraviolet Telescope} ({\it HUT}, Kriss et al. 1992) revealed a number of 
absorption lines from species that span a wide range in ionization potential 
(e.g., O~I to O~VI), as well as fine-structure and metastable absorption lines.
The intrinsic UV absorption was found to be variable in ionic column 
density, but no variations in radial velocities were detected (Bromage et 
al. 1985). The ionization state of the UV absorbers are generally well 
correlated with the UV continuum flux level; the equivalent widths of low 
ionization species are negatively correlated with the flux level, while the 
those of the high ionization species are positively correlated (Bromage et 
al. 1985).
More recent observations of NGC 4151 were obtained with the Goddard 
High Resolution Spectrograph (GHRS) at high spectral resolution ($\sim$15 km 
s$^{-1}$) over limited wavelength regions by Weymann et al. (1997). The GHRS 
spectra revealed that the 
C~IV and Mg~II absorption lines, detected in six major kinematic 
components, were remarkably stable over the time period 1992 -- 1996, with the
notable exception of a transient feature which appeared on the blue
wing of the broad C~IV emission. However, this feature, which was
discussed in our previous paper (Crenshaw et al. 2000b; hereafter Paper I),
turns out to be due to Si~II, rather than a high velocity C~IV component, as
suggested by Weymann et al. (1997), 

X-ray spectra of NGC 4151 reveal the presence of a large intrinsic
column of atomic gas (see, e.g., Barr et al. 1977; Holt et al. 1980).
Based on {\it ASCA} spectra (see George et al.
1998, and references therein), the 0.6 -- 10 keV continuum of NGC 4151 can be fit as a flat power
law ($\alpha$ $=$ 0.3 -- 0.7), with an absorber of column density 
N$_{H}$ $\sim$ 10$^{22 - 23}$ cm$^{-2}$, which is most likely ionized,
as suggested by Yaqoob, Warwick, \& Pounds (1989). 
NGC 4151 is highly variable in X-rays (cf. Perola et al. 1982, 1986), and 
the X-ray absorber appears to respond to changes in the 2 -- 10 keV
flux, although the changes in total column and ionization state
are not simply correlated with the X-ray luminosity (George et al. 1998).
For example, Weaver et al. (1994) found evidence that the ionization state
and column density of the absorber were both higher when the source was
fainter, which may be more easily explained by transverse motion.

In the Hubble Space Telescope ({\it HST}) survey of intrinsic UV absorption 
lines in Seyfert galaxies by Crenshaw et al. (1999), NGC~4151 stood out as an 
unusual object. Of the ten Seyfert galaxies in this sample with intrinsic 
absorption, all showed C~IV and N~V absorption, but only NGC 4151 
showed Mg~II absorption (although, recently, the presence of Mg~II absorption 
has been confirmed in the Seyfert 1 galaxy NGC 3227; Kraemer et al. 2000b).
NGC 4151 is also unusual in the sense that it is the only active galaxy 
known to show metastable C~III$^{*}$ $\lambda$1175 absorption, which is 
indicative of relatively high electron densities ($\sim$ 10$^{9}$ cm$^{-3}$, 
Bromage et al. 1985).

We obtained STIS echelle spectra of the UV spectrum to study the 
intrinsic absorption in detail. These observations are part of a long-term 
project on NGC~4151 by members of the Instrument Definition Team (IDT) of the 
Space Telescope Imaging Spectrograph (STIS) on {\it HST}. In Paper I,
we concentrated on the variability of 
the absorption features in the spectral region surrounding the broad C~IV 
emission. In this paper, we present results on the entire UV spectrum and the 
kinematics and physical conditions in the intrinsic absorbers.

\section{Observational Results}

We observed the nucleus of NGC 4151 with a 0\arcsecpoint2 x 0\arcsecpoint2 
aperture and the STIS E140M and E230M gratings to obtain a spectral coverage of 
1150 -- 3100 \AA\ at a velocity resolution of 7 -- 10 km s$^{-1}$ 
(the ``STIS2'' dataset).
In Paper I, we give the details of the STIS observations and the data reduction 
procedures. We obtained the STIS echelle spectra on 1999 July 19, 
when the continuum and broad emission lines were in a low state compared to 
previous STIS and GHRS observations. We found that the UV continuum decreased by 
a factor of $\sim$4 over the previous two years, which resulted in a dramatic 
increase in the column densities of the broad components of the low-ionization 
absorption lines (e.g., Si~II, Fe~II, and Al~II).  In addition to the absorption 
lines seen in {\it IUE} spectra during previous low states (Bromage et 
al. 1985), we identified numerous broad Fe~II absorption lines that arise from 
metastable levels as high as 4.6 eV above the ground state. We concluded that 
the low-ionization absorption lines are very sensitive to changes in the 
ionizing continuum, and that the extreme variability in these lines could 
explain a couple of puzzling aspects of the UV spectrum of NGC 4151: the 
``transient absorption feature'' in the blue wing of the C~IV emission feature 
(Weymann et al. 1997) and the ``satellite emission lines'' (Ulrich et al. 1985; 
Clavel et al. 1987).

\subsection{Line Identification}

Figure 1 (a -- f) shows the STIS echelle spectra over the entire UV bandpass.
The broad absorption features (FWHM $=$ 435 km s$^{-1}$) appear at a 
heliocentric redshift of 0.0017 (velocity centroid $=$ $-$490 km s$^{-1}$ with 
respect to the nucleus), and correspond to a blend of kinematic components D and 
E in Weymann et al. (1997)
We give our line identifications of the broad absorption features above the
spectrum shown in Figure 1.
Most of the line identifications at the top of Figure 1 agree 
with those claimed in {\it IUE} high-dispersion spectra by Bromage et al. 
(1985), including numerous resonance lines, excited fine-structure lines of 
C~II, O~I, and Si~II, and the metastable C~III$^{*}$ $\lambda$1175 blend. In 
addition we have firm identifications of Ni~II lines with oscillator 
strengths {\it f} $>$ 0.04 (Morton, York, \& Jenkins 1988). Broad Cr~II and 
Mn~II absorption may also be present, but these identifications are less certain 
and may be affected by the numerous Fe~II absorption lines in the spectra (see 
below). We find no evidence in the STIS spectra for the metastable He~II 
$\lambda$1640 and metastable He~I $\lambda$2764, $\lambda$2829, and 
$\lambda$2945 lines claimed by Bromage et al. (1985), although the Fe~II lines 
may be masking these features as well.

Below the STIS echelle spectra in Figure 1, we plot all of the Fe~II UV lines in 
the lists of Silvis \& Bruhweiler (2000), which include multiplets from levels 
as high as 4.6 eV above the ground state. Few regions in the spectra are 
unaffected by Fe~II absorption; this can be seen by noting that local regions of 
the spectra with no Fe~II multiplets plotted below them (and no other absorption 
features above) are much higher than surrounding regions (e.g., 1341 -- 1346 
\AA, 1593 -- 1602 \AA, 2648 -- 2662 \AA, 3014 -- 3044 \AA). It is also clear 
that multiplets from levels high above the ground state must contribute; for 
example, Fe~II UV multiplet HU, arising from a level 4.1 eV above ground, is 
clearly present in the observed spectrum at 1310 -- 1314 \AA~(although the
shorter wavelength lines are blended with Si~II$^{*}$ $\lambda$1309.3)

The narrow kinematic components of absorption identified by Weymann et al. 
(1997) are also seen in a number of ionic species. Figure 2 demonstrates this 
point with plots of the Si~IV $\lambda$1393.8 and S~II $\lambda$1253.8 absorption 
as a function of radial velocity (with respect to the systemic redshift of 
0.00332). For the relatively high ionization line of Si~IV, components A, B 
(Galactic), C, D$+$E (the broad feature), F, and F$'$ are evident (F and F$'$ likely arise in the 
interstellar medium or halo of NGC~4151, Weymann et al. 1997). For the low 
ionization line of S~II, components B, D$+$E, E$'$, and F are present (E$'$ is 
not seen in the high ionization lines). The presence or absence of kinematic 
components in different ions provide valuable information on the physical 
conditions in each component, which will be explored in this paper. An expanded 
atlas of the STIS echelle spectra with our identification of each component of 
absorption is given in Danks et al. (2000).   

\subsection{Saturation and Scattered Light}

There are a number of Galactic (component B) lines that are heavily saturated
(e.g., Si~III $\lambda$1206.5, Si~II $\lambda$1260.4, C~II $\lambda$1334.5, 
Mg~II $\lambda\lambda$2796.3, 2803.5), as evidenced by their extremely small 
residual intensities. The troughs of these Galactic features all lie within 1\% of the 
nearby continuum flux levels, indicating not only heavy saturation, but 
accurate subtraction of the instrumental scattered light as well (see Paper I). 
Although the  residual intensities in the troughs of the broad (D+E) absorption 
lines are much higher ($>$7\% of the continuum in all cases), several clues 
indicate that most of the broad absorption lines are actually highly saturated, 
and the large residual intensities are due to unabsorbed (e.g., scattered) light 
in NGC~4151.

Calculation of the covering factor in the line of sight using the doublet method 
(Hamann et al. 1997; Crenshaw et al. 1999) yields values significantly less 
than unity in the troughs: 0.91 $\pm$ 0.02 for N~V $\lambda\lambda$1238.8, 
1242.8 and 
C~IV$\lambda\lambda$1548.2, 1550.8; 0.85 $\pm$ 0.02 for Si~IV $\lambda\lambda$1393.8, 
1402.8; and 0.62 $\pm$ 0.04 for Mg II $\lambda\lambda$2796.3, 2803.5. This indicates 
unabsorbed emission in the troughs that is not constant as a function of 
wavelength. In addition, the presence of a number of Fe~II lines with small {\it 
gf} values (Silvas \& Bruhweiler 2000), indicates that Fe~II lines with much 
higher values must be saturated. For example, in the 2900 -- 3100 \AA\ region in 
Figure 1f, the Fe~II lines in multiplet 78 with moderate {\it gf} values are 
nearly the same strength as the line with the largest value (at 2990 \AA), and 
even the line with the lowest {\it gf} value in this multiplet is relatively 
strong. Lines in other multiplets with much higher {\it gf} values are clearly 
saturated, although the residual intensities in their troughs can be quite 
large. Finally, the strengths of broad absorption lines with low oscillator 
strengths and/or low abundances (e.g., S~II, Ni~II) indicate that many of the 
other low-ionization absorption lines should be saturated. Thus we have strong 
evidence for heavy saturation of the D$+$E component in most of the absorption 
lines and a significant amount of UV light in the nucleus that is not occulted 
by these broad absorbers.

There are indications that the spectrum of the unocculted light is not simple 
(e.g., a power-law continuum). A power-law continuum drawn through the majority 
of absorption troughs in Figure 1 would lie well above the L$\alpha$, N~V, 
Si~IV, and C~IV absorption troughs, which suggests a lower contribution in these 
regions (e.g., due to absorption features in the unocculted spectrum). On 
the other hand, this continuum would lie well below the level required to account for 
our measured low covering factor for the Mg~II absorption. In addition, at the 
densities required for the presence of the Fe~II metastable absorption 
($>$~10$^{6}$ cm$^{-3}$; Wampler, Chugai, \& Pettijean 1995) or the C~III 
metastable absorption ($>$ 10$^{9}$ cm$^{-3}$; Bromage et al. 1985), the 
Si~II fine-structure line at 1533.4 \AA\ (Figure 1b) should be twice as strong 
as the Si~II resonance line at 1526.7 \AA\ (Flannery, Rybicki, \& Sarazin 1980), 
and it does not appear to be so. The only simple solution to this dilemma is that 
both of these lines are saturated and the flux of unocculted light is greater 
underneath the fine-structure line (i.e, a broad C~IV emission feature in the 
unocculted spectrum).

Thus, the unocculted spectrum must consist of continuum, broad emission lines, 
and broad absorption lines similar to those of the nuclear spectrum itself. The 
most straightforward explanation is that there is a scattering region that is 
outside of the region responsible for the D$+$E component, which reflects a fraction 
of the nuclear spectrum into our line of sight. This is a reasonable conclusion, 
given evidence for electron-scattered radiation in many Seyfert 2 galaxies 
(Moran et al. 2000, and references therein) and the troughs of broad 
absorption-line (BAL) QSOs (Ogle et al. 1999). Additional evidence for electron 
scattering of the nuclear radiation in NGC 4151 comes from ground-based 
spectropolarimetry, which finds significant ($\sim$1\%) polarization in the 
continuum and broad emission lines in relatively large ($\sim$3$''$) apertures 
(Schmidt \& Miller 1980; Axon et al. 1994; Martel 1998). Henceforth, we will 
refer to the unocculted emission in our echelle spectra as ``scattered light''.

Since electron scattering is independent of wavelength, the spectrum of the 
scattered light in NGC~4151 should be the nuclear spectrum averaged over some 
time interval (except for possible velocity shifts or broadening due to 
thermal or bulk motion of the scatterers). However, the extreme variability of 
the UV spectrum of NGC~4151 (Ulrich et al. 1991) and the rather loose limits on 
the location of the scattering region imposed by our aperture size (projected 
distances $<$ 0\arcsecpoint1 or 6.4 pc from the central continuum source) make 
it difficult to determine what 
the time-averaged spectrum should look like. Given that NGC~4151 was in moderate 
to high states for at least 7 years prior to the STIS echelle observations 
(Weymann et al. 1997; Paper I) and evidence that the low-ionization absorption 
lines appear at only low continuum flux levels (Paper I), the most reasonable 
assumption is that the scattered spectrum resembles NGC 4151 in (at least) a
moderate flux state and, hence, contains only the high-ionization 
broad absorption lines.

We have adopted our STIS low-dispersion spectrum, which was obtained in 1998 at 
a high state and covers the entire UV region at high signal-to-noise (Nelson et 
al. 2000a; Paper I), as the basis of our scattered spectrum. In fitting the 
scattered light profile, different scale factors are 
needed for different continuum regions and emission lines. 
For the continuum, we find that the best scale factors increase linearly from 0.12 to 
0.20 over the 1150 -- 3100 \AA\ region. We find scale factors of 0.12 for the 
high-ionization emission lines (e.g., C~IV, N~V, Si~IV) and 0.4 for the 
low-ionization lines (e.g., Mg~II, C~III], the Fe~II emission complex beginning 
at 2300 \AA). Nevertheless, there is 
observational justification for different scale factors. The time-averaged 
nuclear spectrum is likely to reflect a lower state than our high-state 
spectrum, and the variability amplitudes of the low-ionization emission lines 
are much smaller than those of the continuum or high-ionization emission lines 
(Clavel et al. 1990). Also, the variability amplitude of the continuum in 
NGC~4151 (and other Seyfert 1 galaxies) decreases with wavelength in the UV 
(Crenshaw et al. 1996). Thus, larger scale factors are appropriate for the 
low-ionization emission lines and the long-wavelength continuum regions. 

We show our scattered-light spectrum beneath the STIS echelle spectra in Figure 
3 (a -- c, the upper fit to the continuum and broad emission will be discussed 
later). Qualitatively, the scattered-light spectrum provides a good match to 
the troughs of the broad absorption lines. In particular, the scattered spectrum 
no longer falls above the troughs of the high-ionization absorption lines, and 
it matches the troughs of absorption lines embedded in various emission lines 
reasonably well. For example, although the Mg~II absorption appears not to be 
saturated at first glance, the scattered-light spectrum shows that the 
$\lambda$2796.3 line is indeed saturated. Furthermore, tests show that inclusion 
of narrow Mg~II emission in the nuclear spectrum, which should be present at the 
systemic redshift of NGC~4151 (Nelson et al. 2000a), would saturate the Mg~II
$\lambda$2803.5 absorption as well. Although the scattered-light spectrum does 
not provide a perfect match to the troughs of the saturated lines, we conclude 
that that there is a significant flux of scattered light ($\sim$12\% of the 
high-state FUV continuum) within the aperture, and that most of the broad 
absorption lines in the echelle spectra must be highly saturated, indicating 
large ionic columns. In the observed low state, the percentage of light that is
scattered is therefore very high (e.g. $\sim$ 50\% in the long wavelenth
continuum).

\subsection{Absorption Measurements}

In this paper, we focus on the narrow kinematic components A, C, and E$'$, and 
on the broad component D$+$E, since these components are likely to be 
photoionized (directly or indirectly) by the central continuum source.
In addition, we present results on a new component mentioned in Paper I, which 
we will call D$'$; this component is shallow, broad (FWHM $\approx$ 940 km 
s$^{-1}$), highly blueshifted ($-$1680 km s$^{-1}$), and only seen in the high 
ionization lines (N~V, C~IV, and Si~IV). The D$'$ component is not detected in 
previous STIS or GHRS spectra obtained at higher states but it is clearly seen 
as a flattening of the C~IV absorption profile around 1544 \AA\ in the STIS 
echelle spectra (Paper I). Figure 4 shows the D$'$ component in the N~V 
$\lambda$1238.8 absorption. This component was isolated by matching the D$+$E 
component with the profile derived from S~II $\lambda$1253.8 (Figure 2) and then 
subtracting this fit (in optical depth space) from the broad observed blend.

We do not provide results on components B, F, and F$'$ in this paper.
Component B is from our Galaxy, and components F and F$'$ are likely to arise 
from the interstellar medium and/or halo of NGC~4151 (Weymann et al. 1997); 
detailed analyses of these components will be provided in Danks et al. (2000).

To determine the shape of the intrinsic continuum, we used a simple spline fit 
to regions that appeared free of absorption or emission. Similarly, several 
broad emission-line profiles were fit with cubic splines in regions unaffected 
by absorption, and an ``average'' broad profile was adopted (clearly this
is a simplification, since the broad emission lines do not necessarily have the
same profile in individual sources). This profile was then 
reproduced at the known positions of each broad line, and scaled in flux until a
suitable match was obtained. Our resulting estimate of the intrinsic continuum 
plus emission spectrum is given as the upper plot in Figure 3.

Measurements of the narrow absorption components A, C, and E$'$ are straightforward, since 
the absorption lines are relatively unsaturated and resolved (i.e., FWHM $>$ 10 
km s$^{-1}$). Here we assume that these components cover the entire continuum 
plus broad emission regions. Since the residual intensities of the
narrow components are large, a non-unity covering factor would have only a minor
effect on the derived columns. Each absorption-line profile is 
converted to optical depth as a function of radial velocity, and integrated to 
obtain the ionic column density (Crenshaw et al. 1999). Velocity centroids and 
widths (FWHM) are also determined from the optical depth profiles (see Table 1).

Measurements of the broad absorption lines are more complicated.
It is extremely difficult to separate Weymann et al.'s D and E 
components (contrary to our expectations in Paper I), because they are broad and blended together. The velocity centroids 
of all the isolated broad lines are very similar, which indicates that both 
components are present in each line, and it is therefore not possible to isolate 
the D or E component in a particular line. Since any differences in the observed 
profiles of different ions can be explained by saturation effects, contamination 
by other lines (particularly Fe~II), or the scattered light spectrum, 
we can find no evidence that these components are physically distinct. We will 
therefore assume that this is one component (D$+$E). We used a
cubic spline to fit the 
D$+$E component in the S~II $\lambda$1253.8 line, as shown in Figure 2.

Another complication is our finding that most of the broad (D$+$E) lines are 
heavily saturated. Although we have a qualitative idea of what the scattered 
spectrum looks like, we do not know the scattered flux beneath these lines 
exactly, and therefore it is not possible to determine their ionic column 
densities. The only relatively isolated lines that are not clearly saturated are 
S~II $\lambda$1253.8, Ni~II $\lambda$1317.2, and some of the metastable Fe~II 
lines. We determined the velocity centroid and FWHM of D$+$E from these lines, 
and the column densities by integration of their optical depths. To obtain 
absolute lower limits on the D$+$E components of other isolated but saturated 
lines, we used optical depth profiles obtained by assuming no scattered light 
(i.e, a covering factor of 1.0). For doublets that are blended (e.g., C~IV), we 
used the S~II $\lambda$1253.8 optical depth profile to match the two lines 
assuming the proper ratio of oscillator strengths. 

We have retrieved the GHRS spectra of Weymann et al. (1997) from the 
{\it HST} archives to determine the column densities and velocities of the absorption 
components in a consistent fashion. We have also measured the absorption in a 
high-resolution STIS slitless spectrum of the C~IV region observed on 1997
May 25 (Paper I). 
We list our measurements of the velocity centroid and FWHM of 
each component in the echelle spectra in Table 1. These values are consistent with those of Weymann et 
al. (1997), except that his component C is much broader (FWHM $\approx$ 140 km 
s$^{-1}$) than ours. Our own measurements of the GHRS and STIS spectra show no 
evidence for changes in the velocities and widths of any component that are 
larger than those expected from measurement and wavelength calibration errors 
(typically $\pm$10 km s$^{-1}$). We present a comparison of the slitless 
(STIS1), echelle (STIS2), and GHRS (epochs 1 --5)
ionic column densities and lower limits in Table 2.

\section{Modeling The Absorbers}

Photoionization models for this study were generated using the
code CLOUDY90 (Ferland et al. 1998). We have modeled the absorbers
as matter-bounded slabs of atomic gas, irradiated
by the ionizing continuum radiation emitted by the central source.
As per convention, the models 
are parameterized in terms of the ionization parameter,
U, the ratio of the density of photons with energies $\geq$ 13.6 eV
to the number density of hydrogen atoms (n$_{H}$) at the illuminated face of the
slab. Each separate kinematic component was modeled with its
own set of initial conditions, i.e., U, n$_{H}$, and the
physical depth of the slab. As in earlier studies (e.g. Kriss 1998; Crenshaw 
\& Kraemer 1999), a model is deemed successful when the 
predicted ionic column densities provide a good match (i.e., better than
a factor of 2) to those observed. 

The SED for the ionizing radiation in NGC 4151
is difficult to determine, due to the combined effects of intrinsic
absorption in the 
UV (Kriss et al. 1992, 1995) and X-ray (George et al. 1998, and
references therein). Although the narrow emission line ratios are 
sensitive to the SED, Alexander et al. (1999) and Kraemer et al. (2000a)
demonstrated that there is evidence for considerable absorption between the 
central source and the NLR. Hence, while both Alexander et al. and Kraemer et al. 
argued against the presence of a ``Big Blue Bump'' at energies $>$ 13.6 eV,
the narrow lines cannot be used to constrain the SED. As a result, we have assumed the 
simple ``intrinsic'' SED
described in Kraemer et al. (2000a), specifically a series of power-laws of 
the form F$_{\nu}$ $\propto$ $\nu^{-\alpha}$, 
with $\alpha = 1$ below 13.6~eV, $\alpha = 1.4$ over the 
range 13.6~eV $\leq h\nu <$ 1000~eV, and $\alpha = 0.5$ above 1000~eV.

As discussed in Paper I (and references therein), the UV continuum of NGC 4151 
is highly variable.
During the time of these observations, the extinction-corrected 
continuum flux at 1450 \AA~ was f$_{1450}$ $=$ 
1.1 x 10$^{-13}$ ergs cm$^{-2}$ s$^{-1}$ \AA$^{-1}$, assuming an extinction
E$_{B-V}$ $=$ 0.04 (Kriss et al. 1995). This 
is roughly 25\% of the flux of NGC 4151 in a ``high'' state, e.g.
during the GHRS epoch 5 observations (Weymann et al. 1997). In several cases,
we have compared our models of the current (``low'' state) absorbers with the 
high state absorbers observed with the GHRS; in order to do so, we 
increased the ionizing flux by a factor of 4, while, for simplicity, using 
the same SED, although there is some evidence for variations in the latter as a 
function of continuum flux (Edelson et al. 1996).

Since L$\alpha$ is saturated for most of the kinematic components,
the ionization state and effective hydrogen 
column density (N$_{eff}$ $=$ N(H~I) $+$ N(H~II)) of the absorbers is
derived from the ionic columns of the heavy elements\footnote[10]{We
use N(XM) to denote ionic columns, where ``X'' is the
atomic symbol and ``M'' is the ionization state.} 
and, therefore, is sensitive to the elemental abundances. Since there is
no strong evidence for unusual abundances in the nucleus of
NGC 4151 (although see Kraemer et al. 2000a), we have assumed roughly 
solar element abundances (cf. Grevesse \& Anders 1989). They are, by number 
relative to H, as follows: 
He $=$ 0.1, C $=$ 3.4 x 10$^{-4}$, N $=$ 1.2 x 10$^{-4}$, O $=$ 6.8 x 10$^{-4}$,
Ne $=$ 1.1 x 10$^{-4}$, Mg $=$ 3.3 x 10$^{-5}$, Al $=$ 2.96 x 10$^{-6}$,
Si $=$ 3.1 x 10$^{-5}$, P $=$ 3.73 x 10$^{-7}$, S $=$ 1.5 x 10$^{-5}$, Fe $=$ 4.0 x 10$^{-5}$,
and Ni $=$ 1.78 x 10$^{-6}$. We have assumed that the gas is free of
cosmic dust, hence all the elements are fully in the gas phase. 

\section{Comparison of the Observations and Model Results}

\subsection{Component A}

GHRS spectra of Component A revealed 
absorption by C~IV, 
Mg~II, and Si~II (Weymann et al. 1997).
Kriss (1998) was able to match the Mg~II and C~IV columns with a
single-component photoionization model, with U $=$ 0.001 and 
N$_{eff}$ $=$ 2.0 x 10$^{18}$ cm$^{-2}$. The STIS2 data
show absorption from the following ions: H~I, Si~II, Si~III, Si~IV,
C~II, C~IV, and Mg~II. We detected both C~II $\lambda$1334 and 
C~II$^{*}$ $\lambda$1336; the ratio of N(C~II)/N(C~II$^{*}$) 
is $\approx$ 1.2, indicating that n$_{e}$ (electron density) 
is $\sim$ 100 cm$^{-2}$ (Srianand \& Petitjean 1999), which is 
consistent with the N(Si~II)/N(Si~II$^{*}$) ratio, within the errors. After Shields \& Hamann (1997), 
the C~IV recombination timescale is $t$
$\sim$ (n$_{e}$ N(C~IV)/N(C~III) $\alpha$$_{C~IV}$)$^{-1}$. Using 
$\alpha$$_{CIV}$ (the 
radiative recombination rate) from Shull \& van Steenberg (1982), 
$t$ is $\sim$ 800 yrs.
We measured a C~IV column of (4.1 $\pm$ 0.3) x 10$^{13}$ cm$^{-2}$, which is close to the earlier GHRS 
results (see Table 2) as expected, given the 
long recombination time. The Mg~II and Si~II columns are identical to those
from GHRS epoch 5, within the measurement errors.

We were able to fit the ionic columns densities of A in the 
STIS2 spectra with a single component 
photoionization mode assuming U $=$ 0.0012, and N$_{eff}$ $=$ 
1.3 x 10$^{18}$ cm$^{-2}$, similar to Kriss (1998). The model
predictions are compared to the measured columns in Table 3 (note that
the model predicts total ionic columns, rather than separate predictions for 
ground and excited states of ions). 
The predictions matched the measured columns quite well, although N(Si~III) is 
overpredicted, which may be 
due to the difficulty in determining the profile of the blue wing of
L$\alpha$. 

As shown in Table 2, between GHRS epochs 4 and 5 there was a significant 
drop in the Mg~II and Si~II columns. Given the low densities derived for 
Component A, these changes are most likely the result of transverse motion, 
rather than a response to changes in the ionizing continuum. 
We were able to match the GHRS epoch 4 results by adding a second component,
with U $=$ 0.0003 and N$_{eff}$ $=$ 4 x 10$^{17}$ cm$^{-2}$; assuming the
two components are at the same radial distance, the additional component would 
have a density of $\sim$ 400 cm$^{-3}$. The additional low ionization gas contributes no 
C~IV absorption, as required by the lack of variation of N(C~IV).
This component must have a transverse velocity of $\sim$ 2000 km s$^{-1}$ in
order to pass out of our line of sight to the broad line region (BLR) within 1.5 years, 
assuming the BLR of NGC 4151 is 4 light days in diameter (Clavel et al. 1990); this is not
unreasonable, considering that the radial velocity of Component A is
$\sim$ 1588 km s$^{-1}$.

\subsection{Component C}

Both C~IV and Mg~II absorption lines were identified with Component C 
in the GHRS spectra (Weymann et al. 1997). 
Based on these ionic columns, Kriss (1998) modeled the absorber as a single slab
with U $=$ 0.002 and N$_{eff}$ $=$ 1.25 x 10$^{18}$ cm$^{-2}$.
Our re-examination of the GHRS spectra revealed the presence of 
absorption by Si~II $\lambda$1526 in Epoch 1. 
In addition to these ions, the STIS2 spectra show absorption from 
Si~III, Si~IV, and C~II. Based an the upper limit
for C~II$^{*}$ $\lambda$1336, N(C~II$^{*}$)/N(C~II) is $\sim$ 0.2,
which indicates n$_{e}$ $\sim$ 10 cm$^{-2}$. At such low densities, Si~II$^{*}$
$\lambda$1533 should be weak or absent (Flannery et al. 1980), which is 
indeed the case. Also, the ionic columns should not respond to the changes in
ionizing flux, which is true for Mg~II, Si~II, and C~IV (Table 2).

We modeled Component C as a single slab, with U$=$ 0.0012, and N$_{eff}$
$=$ 1.0 x 10$^{18}$ cm$^{-2}$. The model predictions 
and measured ionic columns for are listed in Table 4. Most of the ionic
columns are well fit, albeit with overpredictions of Si~III (see above) 
and Si~II.  

\subsection{Component D$+$E}

As noted in Section 2, we have not tried to deconvolve the Weymann et al.'s
Components D and E. The cumulative absorption component (D$+$E) possesses 
strong 
high ionization lines, e.g. N~V, C~IV, and Si~IV. In our STIS2 
spectra, a number of lower ionization
absorption lines are associated with Component D$+$E, the most striking of 
which are those from metastable levels of Fe~II (Paper I), including those 
with lower levels up to 4.1 eV above the ground state. With the resolution of 
the STIS echelle, we have now established that the C~III$^{*}$ $\lambda$1175 
absorption, which arises from a level 6.5 eV above the ground state, is also 
associated with D$+$E, indicating densities of n$_{e}$ $\geq$ 10$^{9}$ cm$^{-3}$.

The population of metastable states of C~III and Fe~II
likely result from collisional excitation (although, for a discussion
of the excitation of Fe~II metastable states by UV radiation, see Verner et al. 
[1999]). The relative populations
of the C~III metastable, $^{3}$P$^{o}$, and 
ground, $^{1}$S$_{L}$ states, as a function of n$_{e}$ 
and temperature, have been calculated by Bromage et al. (1985) and Kriss et 
al. (1992). For a gas characterized by n$_{e}$ $=$
10$^{9.5}$ cm$^{-3}$ and T $=$ 2 x 10$^{4}$K, a few percent of the 
C~III will be in the excited state (while collision
rates for temperatures below 2 x 10$^{4}$ K are unavailable at the
time of this writing, it is probable that the relative population of
the $^{3}$P$^{o}$ state will be smaller at lower temperatures). Under such conditions, 
{\it at least} the lowest 25 energy levels for Fe~II ($\leq$ 2.64 
eV above
the ground state) will be in statistical equilibrium (Verner et al. 1999).
The ratio of Si~II$^{*}$/Si~II cannot be used as a density indicator,
since it is clear that these lines are saturated (see Section 2).

As discussed in Section 2, the measurement of the ionic column densities
for Component D$+$E is hampered by the complex profile of the underlying
scattered component and the fact that the lines from the most
abundant ions are saturated. The only isolated unsaturated lines are
are S~II $\lambda$1254 (which has a weak oscillator strength, $f$ $=$ 0.01; 
Morton et al. 1988), Ni~II $\lambda$1317 (which has a low abundance;
Section3), and some of the weaker
Fe~II metastable lines. We have, therefore, used the S~II and Ni~II columns to
constrain the fraction of low ionization gas in the absorber. Although there
are no weak high ionization lines in the STIS data, 
P~V $\lambda\lambda$1117.98, 1128.01 has been detected in 
HUT Astro-1 and 
Astro-2 observations (Kriss et al. 1992; Kriss et al. 1995) and with the Berkeley
sepctrometer aboard {\it Orbiting and Retrievable Far and Extreme
Ultraviolet Spectrometers (ORFEUS) - Shuttle Pallet Satellite (SPAS) II} mission
(Espey et al. 1998). Based on the
equivalent widths determined by Espey et al., and the appearance that
the lines are unsaturated, N(P~V) is $\approx$
1.2 x 10$^{14}$ cm$^{-2}$. Since the ionization potentials of P~IV and C~III
are similar (51.4 eV and 47.9 eV, respectively; Allen 1973), it is reasonable
to expect that the ratio of ionic columns of P~V and C~IV will be 
approximately the same as that of the elemental abundances of phosphorus and 
carbon (i.e, 0.001; cf. Grevesse \& Anders 1989). Hence, at 
the time of {\it ORFEUS-SPAS II} observations, during which NGC 4151 was in a 
high flux state, N(C~IV) must have been $\sim$ 10$^{17}$ cm$^{-2}$. 
Although there are no far-UV observations contemporaneous with the STIS echelle
observations, we will show that it is plausible that the columns of C~IV and P~V have remained
relatively constant since the time of the {\it ORFEUS-SPAS II} mission. 

Based on these assumptions, we have modeled the Component D$+$E as a single 
slab, with U $=$ 0.015, n$_{H}$ $=$ 3.2 x 10$^{9}$ cm$^{-2}$, and N$_{eff}$ $=$ 2.75 x 10$^{21}$ cm$^{-3}$. The
predicted and measured ionic columns are listed in Table 5. We ran two models,
one assuming thermal broadening and the other a microturbulence velocity of
435 km s$^{-1}$, and found no significant differences in the predicted
ionic column densities (the results we discuss are from the former model).
The model is successful in reproducing the Ni~II and S~II columns, within
the measurement errors. In all other cases, with the exception of Al~III, the
model predictions exceed the lower limits for the ionic column densities, which
is a {\it necessary, but insufficient test} of the model. 
The predicted average electron temperature is $\sim$ 1.6 x 10$^{4}$K, hence our
assumption that fractional population of the $^{3}$P$^{o}$ state of C~III is 
$\sim$ 1\% is reasonable. The 
predicted N(Al~III) is close to the derived upper limit, which may indicate
that the Al~III lines are not highly saturated, or that the 
aluminum may exceed solar, as has been suggested for BALQSOs (Shields 1997).

As an additional test of the model, we ran a high state version, increasing
U by a factor of 4 (see above), while holding N$_{eff}$ fixed. The results
are listed in Table 5, alongside the low state model predictions. Notably,
the Fe~II and Si~II columns are small enough that their absorption lines
would be weak or below the limit of detectability, which is precisely the
case for the STIS1 (Paper I) and GHRS (Weymann et al. 1997) observations, which
were all obtained at a high state. On the other hand, Mg~II
absorption would still be present; the predicted column density is
similar to that seen in GHRS epoch 5 (which we determined to be 
$\approx$ 1.0 x 10$^{14}$ cm$^{-2}$). The higher ionization lines,
such as N~V $\lambda\lambda$1238, 1241 and C~IV $\lambda\lambda$ 1548, 1551,
would remain heavily saturated, and hence do not change appreciably
in profile, as is the case (see Paper I). Most important, the C~IV column 
density does not change significantly, while N(P~V) is close to the value 
derived from the {\it ORFEUS - SPAS II} observations. Therefore, given this
model solution, it is possible to use the high state value of N(P~V) to 
constrain N$_{eff}$. Although there may be other possible solutions, we will
show in section 4.4 that the physical conditions in Component
D$^{'}$ depend strongly on the column density and ionization state of
Component D$+$E, and hence provide a further test.

\subsection{Component D$^{'}$}

The appearance of Component D$^{'}$ was discussed in Section 2.3, and, as noted,
it is present in N~V, C~IV, and Si~IV (see Table 6). 
Since N(Si~IV) is $=$ 4.5 x 10$^{14}$ cm$^{-2}$, there must be a significant
column of C~III, although no corresponding C~III$^{*}$ $\lambda$1175 is
detected, indicating that the electron density is $<$ 10$^{9.5}$cm$^{-3}$. 
Assuming that this component is 
directly ionized, we were
unable to fit the measured ionic columns with a single-component model. Our best
fit was a two component model, with the C~IV and Si~IV absorption arising
in one component (U $=$ 0.001; N$_{eff}$ $=$ 
2 x 10$^{20}$ cm$^{-2}$), and the N~V arising in more highly ionized gas
(U $\geq$ 0.1, N$_{eff}$ $\geq$ 1 x 10$^{20}$ cm$^{-2}$). However,
if the ionizing flux is increased by a factor of 4, the C~IV column
predicted by the lower ionization component {\it increases}, due to the
ionization of the large C~III column; hence a
directly ionized, multi-component model does not predict the absence
of Component D$^{'}$ while NGC 4151 is in a high state.

The coincidence of the appearance of D$^{'}$ at the same time as
the low ionization lines from D$+$E can be best explained if the former
is screened by the latter. To demonstrate this, we generated a single-component
model, with N$_{eff}$ $=$ 1 x 10$^{20}$ cm$^{-2}$, using the ionizing 
continuum filtered by D$+$E as the input spectrum (see
Figure 5a). In order to reproduce the observed high ionization lines, 
assuming that these components are at the same approximate
radial distance, the D$^{'}$ component must have U $=$ 0.012, which yields
a density of n$_{H}$ $=$ 2.6 x 10$^{7}$ cm$^{-3}$. As shown
in Table 6, the model predictions provide a good fit to the N~V, C~IV, and 
Si~IV columns, and indicate that Si~II and Mg~II lines should be
weak or absent (due to the small column densities and broadness of
this feature). A second model was generated, using the filtered continuum
from the high state model for Component D$+$E as the input (Figure 5b), 
while 
holding n$_{H}$ and N$_{eff}$ fixed, which results in U $=$ 4.15. The results are listed in the 3rd column of Table 6 
and, as required by the observations, the predicted column densities are quite 
small. Therefore, the appearance of D$^{'}$ while NGC 4151 was in its low 
state is the direct result of the higher EUV opacity of D$+$E .

\subsection{Component E$^{'}$}

Component E$^{'}$, isolated kinematically by Weymann et al. (1997), is
a low ionization absorber, as evidenced by the presence of Mg~II $\lambda$2803
and absence of C~IV $\lambda\lambda$ 1548,1551. Based on these constraints 
Kriss (1998) determined the ionization parameter for this component to be
U $<$ 0.0004, with N$_{eff}$ $\sim$ 10$^{17}$ -- 10$^{18}$ cm$^{-2}$. 
In addition to Mg~II, we have found absorption lines of O~I, C~II, Si~II,
and Fe~II associated with this component, but, as with the GHRS results,
no high ionization lines (Table 7). Interestingly, all of the ionic species
present arise from states with ionization potentials below 13.6 eV. 
Among the more interesting features is the presence of absorption by
the fine structure line O~I$^{*}$ $\lambda$1304 (in addition to the
ground state line O~I $\lambda$1302). The ratio of N(O~I$^{*}$)/N(O~I) is
$\approx$ 0.64, which is close to the ratio of the statistical weights of the
two levels, indicating that they are in collisional equilibrium. Based on the
transition probabilities and collision strengths (Pradhan \& Peng 1995, and 
references therein), Component E$^{'}$ must have 
n$_{e}$ $\geq$ 10$^{6}$ cm$^{-3}$. This is consistent with the density
indicated by the N(Si~II$^{*}$)/N(Si~II) ratio, $\sim$ 2, within the 
measurement errors, requiring n$_{e}$ $>$ 10$^{3}$ (Flannery et al. 1980).

We modeled E$^{'}$ as a single slab, with U $=$ 0.0001, n$_{H}$ $=$ 1 x 
10$^{6}$ cm$^{-3}$, and N$_{eff}$ $=$ 1.5 x 10$^{18}$ cm$^{-2}$. The 
model predictions are listed in Table 7, and show good agreement with the
measured ionic columns. The predicted Fe~II column is greater than that 
measured, which is evidence for a substantial population in
metastable states, as one would expect at this density (Wampler et al. 
1995) (the metastable Fe~II lines were not detected in our spectra, due
to their substantially lower oscillator strengths). A high state comparison model was run, with U increased by
a factor of 4. The predictions for N(Mg~II) and N(Si~II) were 6.7 x 10$^{12}$ 
cm$^{-2}$ and 1.3 x 10$^{13}$ cm$^{-2}$, respectively, in good agreement with the
GHRS epoch 4 values (see Table 2). This indicates the 
increase in ionic column density of the low ionization lines between
GHRS4 and STIS2 is likely due to a decrease in the ionizing flux.
The columns for the high ionization lines remain small, e.g.,
N(C~IV) $=$ 7.4 x 10$^{12}$ cm$^{-2}$, which would make them difficult to
detect.

\section{Discussion}

\subsection{The Geometry of the Circumnuclear Gas}

The presence of absorption from both ground and fine structure absorption
lines of O~I, Si~II , and C~II in several kinematic component, and the 
absorption from metastable C~III and Fe~II in Component D$+$E, allows us to 
constrain the
densities of the absorbers. Combined with the evidence for screening of Component
D$^{'}$ by Component D$+$E, and a significant contribution from
an unocculted scattering region, we can begin to define the structure
of the circumnuclear gas in NGC 4151. 

Based on the flux near the Lyman limit and our assumed SED, the number of ionizing photons emitted by 
the central source is Q $=$ 2 x 10$^{53}$ s$^{-1}$. From our determination
of their densities\footnote[11]{Note that the densities for Components
D$+$E and E${'}$ are lower limits.}
and ionization parameters, we place the absorbers at the 
following radial distances: Component C, 2.15 kpc; Component A, 681 pc; 
Component E$^{'}$, 23.6 pc, and Component D$+$E, 0.03 pc. 
As noted in the previous sections, we have assumed that Components A, C, and 
E$^{'}$ are directly ionized by the central source, and hence not screened by
intervening gas. However, it is entirely plausible that A and C are
screened by Component D$+$E. To test this, we generated models for these 
components using
the high state filtered continuum (Figure 5), and found little change in the
resulting column densities. 
The main effect is that these components would lie at $\sim$ 2/3 the radial
distance determined from the direct ionization models. Due to its
greater density, Component E$^{'}$
must respond to short timescale flux and opacity changes, and we were 
unable to model the small oberved changes in the ionic columns by assuming 
that this component is screened
by D$+$E. Therefore, we suggest that E$^{'}$ is directly ionized, but
not in our line-of-sight to the BLR and central source and, instead, is
only occulting scattering emission\footnote[12]{If this is the case then
the changes in the measured ionic columns over the last several years
could at least partially be a result of an increased contribution of the scattered continuum
to the total UV flux.} (which contributes 1/3 -- 1/2 of the
UV flux in the STIS2 data; see Section 2.2). Based on our model results, the absorbers decrease in density with increasing radial 
distance, faster than distance$^{-2}$ up to Component A.
The relative positions of the absorbers and the scattering region are shown in Figure 6. 

\subsection{The Scatterer}

Based on our analysis of the scattered profile, we can begin to constrain the
physical characteristics of the scatterer. Assuming isotropic scattering by free
electrons (see Section 2.2), for 
small electron scattering optical 
depths ($\tau$ $<$ 1), the reflected fraction of continuum radiation is
$\approx$ N$_{e}$F$_{c}$$\sigma$$_{thompson}$ (the product of the column
density of free electrons, the covering factor, and the Thompson cross-section,
respectively). In order to scatter $\sim$ 15\% of the UV radiation (see Section 2),
the scatterer must have N$_{e}$ $\sim$ 2 x 10$^{23}$ cm$^{-2}$ (for a covering
factor of unity). From STIS G430M spectra taken on
May 15, 2000, we determined that the total broad $+$ narrow H$\beta$ 
flux from the central 0\arcsecpoint1 was $\approx$ 5.8 x 10$^{-12}$ 
ergs s$^{-1}$ cm$^{-2}$ (Nelson et al.
2000b), or L$_{H\beta}$ $\approx$ 1.2 x 10$^{41}$ ergs s$^{-1}$ 
(the distance to NGC 4151 is 13.3 Mpc, 
assuming H$_{o}$ $=$ 75 km s$^{-1}$ Mpc$^{-1}$). Since reverberation
mapping indicates that the broad-line emission arises in relatively dense
gas (cf. Clavel et al. 1990), which will not contribute large
columns of free electrons, we may assume that the
scatterer makes only a small contribution (i.e., $<$ 10\%) to the 
broad H$\beta$ emission. Since the total H$\beta$ emission depends
on the number of recombinations per unit volume, this provides an upper limit 
to the density. Assuming Component D$+$E is $\approx$ 0.03 pc
from the nucleus, and does not cover the scattering region, 
we may assume that the scatterer is coincident or lies further from the 
nucleus, e.g., 0.1 pc. Finally, this component cannot
be opaque to the ionizing radiation, since there is strong evidence
that the NLR is powered by the central source (Alexander et al. 1999; 
Nelson et al. 2000a; Kraemer et al.
2000a). With these constraints, the scatterer can be characterized with the
following parameters\footnote[13]{The temperature at the 
illuminated face of this component is 2.2 x 10$^{5}$K, hence it is not
thermally stable (Krolik, McKee, \& Tarter 1981)}: U $\geq$ 4.5, N$_{eff}$ $=$ 2 x 10$^{23}$ cm$^{-2}$, and 
n$_{H}$ $\leq$ 1.0 x 10$^{6}$ cm$^{-3}$. Recent {\it Chandra} High
Energy Transmission Grating spectra of NGC 4151 reveal
an extended ($\sim$ 1.6 Kpc), X-ray emission line region (Ogle et al. 2000),
comprised of a two-phased medium: a hot (10$^{7}$K), collisionally
ionized plasma, in which high ionization lines such as Fe XXV 1.79 \AA~ arise,
and photoionized gas at temperatures $\leq$ 4 x 10$^{4}$K, which give rise to
emission lines and radiative recombination continua from lower ionization 
species. It is possible that scatterer is part of the 
hot medium, although Ogle et al. find
little scattered continuum below 3 keV.

\subsection{The X-ray absorber}

The scatterer model predicts large O~VII, O~VIII, Ne~IX, and Ne~X columns
(9.7 x 10$^{19}$ cm$^{-2}$, 2.3 x 10$^{19}$ cm$^{-2}$, 1.7 x 10$^{19}$
cm$^{-2}$, and 3.5 x 10$^{18}$ cm$^{-2}$, respectively),
and, hence would appear as a ``warm absorber'' if viewed in our line-of-sight.
This leads to the question whether the X-ray absorber and the
scatterer are the same component, as suggested by Krolik \& Kriss (1995)
for Seyfert galaxies in general. This 
is particularly interesting since
none of the UV components possesses sufficient X-ray opacity to account for the
observed X-ray absorption. For example, Component D$+$E has the greatest 
column density
and highest ionization parameter, but it transmits a considerable fraction of
the intrinsic continuum at energies $>$ 300 eV (see Figures 5a and 5b), which 
is clearly not
observed in NGC 4151 (George et al. 1998). Furthermore, as noted above,
the scatterer must have a high covering factor, and hence, it is likely that
it covers the central source. To address this, we
examined the 
{\it ASCA} observations of NGC 4151 from 1993 November 5. Since NGC 4151
was in a higher flux state at this time than during the STIS2 observations, we can
assume that the scatterer was in at least as high a state of ionization
as the one we have modeled. However, the predicted
ratios of N(O~VIII)/N(O~VII) and N(Ne~X)/N(Ne~IX) are too large to fit
the aborption profile in the {\it ASCA} data, and the fit cannot
be improved by varying the fraction of scattered light, column density of the
scatterer, or the X-ray photon index
($\Gamma$).
Instead, the best fit parameters are as follows:
U $=$ 0.86, N$_{eff}$ = 4 x 10$^{22}$ cm$^{-2}$, $\Gamma$ = 1.5, and a
scattered fraction of $\approx$ 0.045 of the intrinsic
flux (for the details of the analysis
and fitting statistics, 
see George et al. 1998). The predicted ionic column densities for the
X-ray absorber are listed in Table 8. The model
predicts large C~IV and N~V columns, and hence could explain some
of the UV absorption. Nevertheless, N(P~V) and N(Mg~II) are too low, even
for a high state period, hence we believe that this component in not
responsible for the the bulk of the UV absorption. Furthermore, the C~IV and N~V
absorption from the X-ray absorber could be buried in the deep D$+$E 
absorption trough, and hence undetectable, as suggested by Kriss et al. (1992). In summary,
in addition to the UV absorbers and highly ionized scatterer, there
is a third component of gas along our line-of-sight with an 
intermediate ionization parameter that produces the
bulk of the X-ray absorption, similar to the multi-phased NLR suggested
by Ogle et al. (2000). 

The decoupling of the X-ray and UV absorption presents a problem: if the
absorbers occult the same continuum source, they must screen one another.
We ran a series of photoionization models to test
the effects of screening. If the X-ray absorber is screened by the Component
D$+$E, we found it will drop significantly in ionization as the continuum flux
drops and the screening gas becomes increasingly opaque to the ionizing
radiation, similar to the case for Component D$^{'}$. However, we
find no evidence of the huge low ionization columns (e.g. N(Ni~II) $\geq$ 10$^{16}$
cm$^{-2}$) which would arise from the X-ray absorber in such a situation. Hence, in this
scenario, the X-ray absorber must be of such low density that it does
not respond on timescales of several years (i.e., 
n$_{H}$ $<$ 10$^{3}$ cm$^{-3}$) which, in turn, requires that the short 
timescale variations in the X-ray absorption (Weaver et al. 1994; 
George et al. 1998) are due solely to transverse motion. 
On the other hand, if the X-ray absorber screens Component D$+$E from the 
ionizing source, as the ionizing
flux dropped, we found that the UV absorber becomes far too neutral, unless the
columns density of the X-ray absorber decreases by at least a factor of 4.

Although we cannot rule out the possibility that the UV and X-ray absorbers
screen one another, we suggest another scenario. If the regions 
producing the UV and X-ray continua are not equally extended, each
region could be covered by the respective absorber, i.e., the UV aborber
covers the UV source and the BLR clouds, while the X-ray absorber covers
only the X-ray source. Since the UV absorber must cover the BLR ($\sim$
4 light days; Clavel et al. 1990), the
simplest picture would have the X-ray absorber closer to the central
source, occulting a smaller X-ray emitting region. The problem with this
scenario is that the fraction of scattered light is apparently smaller in the 
X-ray than in the UV ($\sim$ 0.04 versus $\sim$ 0.15 at a high state).
However, this could be the result of 1) a (relatively) anisotropic
X-ray continuum, as compared to the UV or 2) an isotropic X-ray absorber inside
the radius of the scattering region, which modifies the X-ray continuum before
it reaches the scattering region. In the latter case, a larger
scattered light contribution would be necessary to fit the {\it ASCA} data.
Since our fit to the UV data requires the presence of absorption profiles
in the scattered light, and there is evidence for the presence of an
X-ray absorber inside the NLR of NGC 4151 (Kraemer et al. 2000a), it would
not be surprising that the scattered X-ray continuum would show strong
absorption features. 

\section{Summary}

We have used medium resolution echelle spectra obtained with {\it HST}/STIS
to study the physical conditions in the intrinsic UV absorbers in
the Seyfert 1 galaxy, NGC 4151, while it was in a recent low continuum state. 
The present paper is a detailed
study following our observational results from Paper I. We have
determined the following regarding the circumnuclear gas in NGC 4151.

1. Each of the kinematic components can be modeled as a single slab
of atomic gas, of roughly solar elemental abundances, ionized by the
EUV -- X-ray continuum radiation emitted by the central source. The combined
kinematic component D$+$E, which we have not attempted to
deconvolve, has the largest column density (2.75 x 10$^{21}$ cm$^{-2}$) and
ionization parameter (U $=$ 0.015) of the absorbers. The C~III and
Fe~II metastable lines arise in this component, indicating it is 
also the densest (n$_{H}$ $\geq$ 3 x 10$^{9}$ cm$^{-3}$) and closest to the
ionizing source. Component D$^{'}$, which was not detected in earlier high
state spectra, appears to be screened from the nuclear source by
Component D$+$E; it becomes too highly ionized to detect as the opacity
of D$+$E decreases when NGC 4151 enters a high flux state. Our model
predictions match the observed changes in the ionic columns; specifically that
the denser components D$+$E, D$^{'}$, and E$^{'}$
responded to the decrease in continuum flux since the time of the GHRS
spectra, while the low density components A and C, have not (as a result of 
their long recombination timescales). None of the UV absorbers has sufficient
X-ray opacity to produce the observed X-ray absorption.

2. We have been able to constrain the densities of the UV absorbers 
based on the ratios of fine structure to
ground state C~II, O~I, and Si~II lines, and the presence of lines from
metastable states of C~III and Fe~II. From these densities, and the ionization
parameters determined from the fits to the ionic column densities, we can
determine the relative radial positions of the individual absorbers, as
shown in Figure 6. Based on these results, it is apparent that the
UV absorbers extend over a large range in radial distances, and 
decrease in density with distance.

Weymann et al. (1997) found that the radial velocities of Components A and C had
changed little between GHRS epochs 2 and 5. From this, they determined
the limit for steady radial acceleration for Component A to be 
$\sim$ 1 x 10$^{-3}$ cm s$^{-2}$. Assuming that the cloud is optically thin,
and a mean observed continuum flux for NGC 4151, this limit requires that 
Component A lies at a radial distance of $\geq$ 0.7 pc. Since the
distances we have determined for Components A, C, and E${'}$ are all 
significantly greater than this limit, we find no apparent conflicts
with the Weymann et al.'s conclusions. Since Component D$+$E lies much closer
to the nuclear source, one might expect that it should experience
a noticeable acceleration. However, since D$+$E is optically thick in the
Lyman continuum and has a large column density, it is probable that 
radiative acceleration will be less efficient (Williams 1972; Mathews 1974). 
Additionally, the evidence for transverse motion in the UV and X-ray 
is evidence for more complex dynamics, which we intend to 
address in a subsequent paper.
 
3. Based on the residual flux in the troughs of staturated absorption lines
associated with Component D$+$E, we estimate that at least 15\% of the 
high-state continuum emission, broad emission lines, and absorption lines are scattered
into our line of sight, from a plasma that extends outside the solid
angle subtended by the absorber (in Figure 6 we show a possible
location for the scatterer). For a covering factor of unity, the scatterer must have a free electron 
column density of N$_{eff}$ $\sim$ 2 x 10$^{23}$ cm$^{-2}$. Based on the constraints
on the total H$\beta$ flux, the covering factor, and the EUV opacity of the 
scatterer, we argue that it must be highly ionized (U $\geq$ 4.5) and, hence,
cannot produce the observed X-ray absorption. Instead, a smaller
column of lower ionization gas (N$_{eff}$ = 4 x 10$^{22}$ cm$^{-2}$,
U $=$ 0.86) must also be present. There are several possibilities
regarding the relative position of this
additional component with respect to the UV absorbers and scatterer. Also,
the reason for the apparent differences in the fractions of UV and X-ray 
scattered light remains an open question, which can be better addressed 
through simultaneous UV ({\it HST}/STIS) and 
X-ray ({\it Chandra}) observations of NGC 4151.

\acknowledgments

 S.B.K. and D.M.C. acknowledge support from NASA grant NAG5-4103. We thank
Fred Bruhweiler, Tahir Yaqoob, and Jane Turner for enlightening discussions.

\clearpage

\figcaption[fig1.ps]{STIS echelle spectra of the nucleus of NGC 4151; the 
spectra have been smoothed by a boxcar function of 7 pixels for display 
purposes. Identifications of the broad absorption lines
associated with D$+$E, except for Fe~II, are 
given above the spectra. Fe~II multiplets are plotted below the spectra, with 
the length of each vertical line representing the {\it gf} value from Silvas \& 
Bruhweiler (2000).}

\figcaption[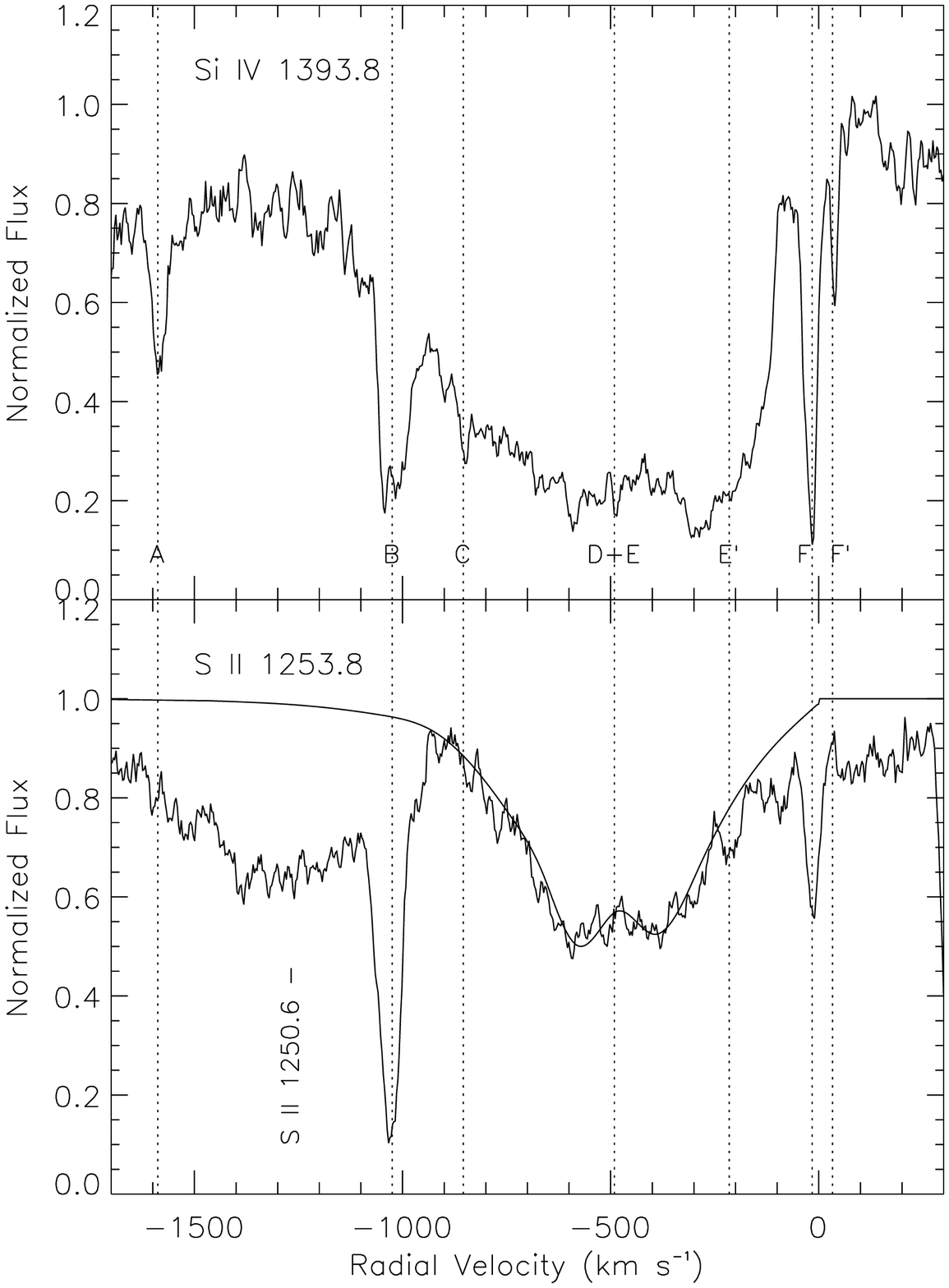]{Plots of the Si~IV $\lambda$1393.8 and S~II 
$\lambda$1253.8 lines as functions of radial velocity (with respect to the 
nucleus of NGC~4151 at a redshift of 0.00332). The spectra have been normalized 
by dividing by the continuum plus emission fit (Figure 3). The expected 
positions of the absorption components are plotted as vertical dotted lines.
The smooth line in the lower plot is our fit to the D$+$E component of S~II;
note there may be a slight contamination at $-$100 to $+$300 km s$^{-1}$ due to 
an Fe II line from multiplet UV CW. Component B is Galactic, and
Components F and F$^{'}$ are most likely from the interstellar medium or
halo of NGC 4151.}

\figcaption[fig3.ps]{STIS echelle spectra of NGC~4151. The lower plots give our 
estimate of the scattered light spectrum (see text). The upper plots give our 
continuum plus broad-emission fits to the echelle spectra.}

\figcaption[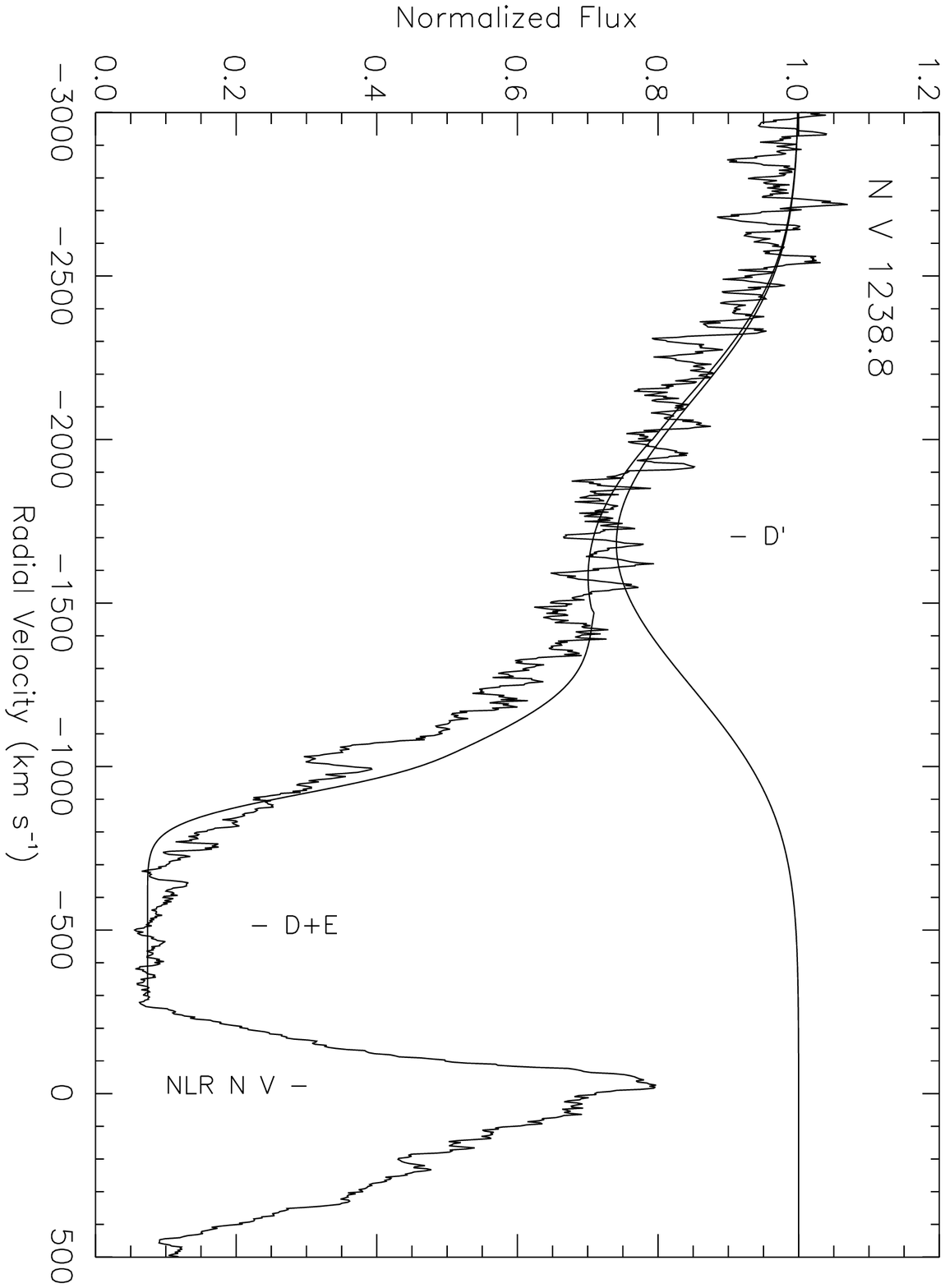]{Plot of the N~V $\lambda$1238.8 absorption as a function of 
radial velocity. The upper smooth plot gives our fit to the D$'$ component, the 
lower smooth plot gives the sum of the D$'$ and D$+$E components. The location 
of N~V emission from the narrow-line region (NLR) is noted. The scattered
component has not been subtracted in this plot.}

\figcaption[fig5.ps]{The incident (solid line) and transmitted (dotted line)
continua for the models of Component D$+$E: a) the
low flux state model and b) the high flux state
model (see text, Section 4.4). The flux values for the incident
continuum are in log($\nu$F$_{\nu}$) at the face of the slab (in 
units of ergs s$^{-1}$ cm$^{-2}$).}

\figcaption[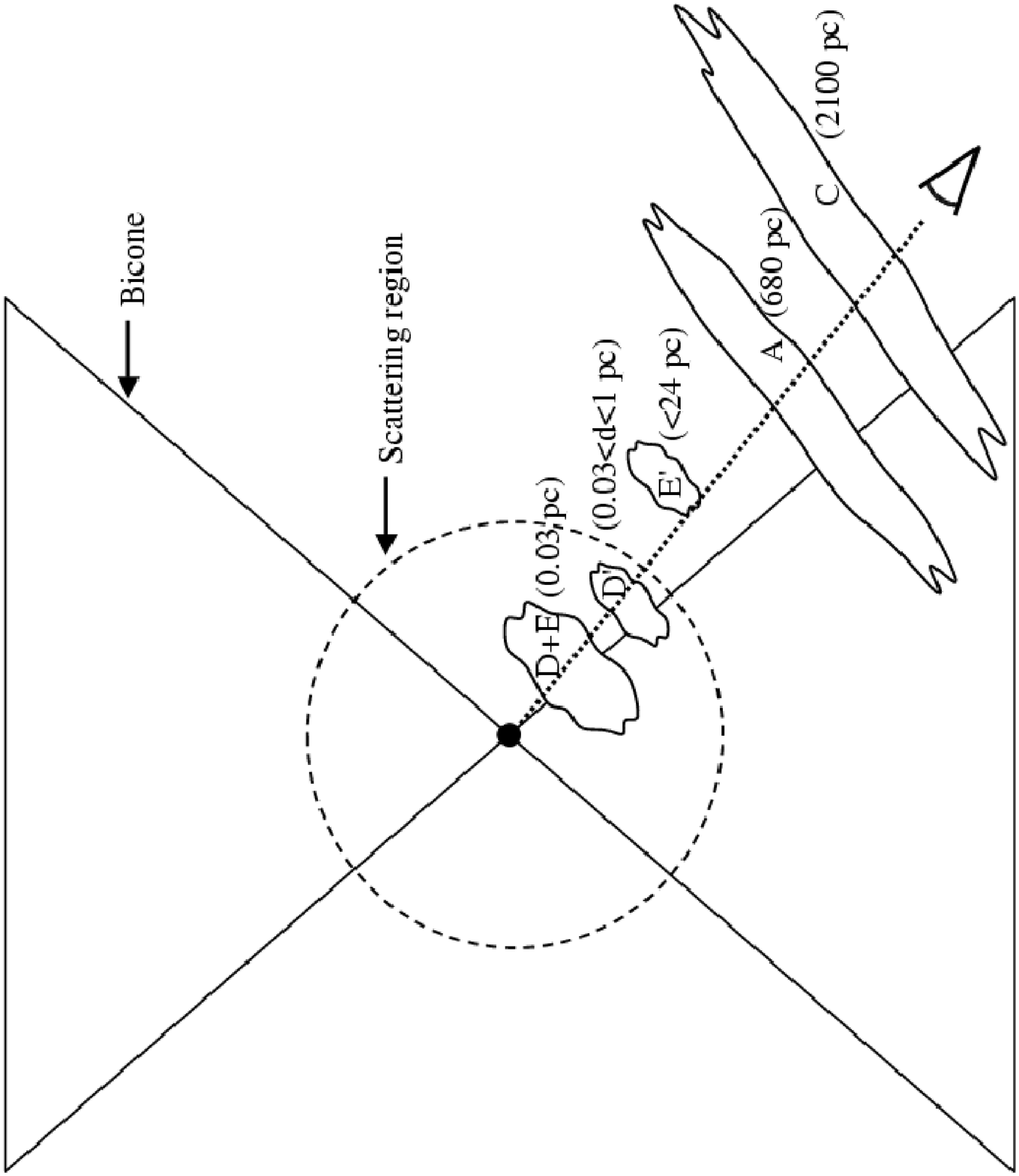]{Schematic diagram of the nuclear region of NGC 4151,
showing the relative positions of the central source, the emission-line
bicone, and the UV absorbers. The viewing angle relative to the bicone
is given by the kinematic model by Crenshaw et al. (2000a).
The dotted line represents a possible
radial position and distribution of the scattering medium, although
the only real constraint is that it is not completely covered by D$+$E
in projection.}

\clearpage
\begin{deluxetable}{ccc}
\tablecolumns{3}
\footnotesize
\tablecaption{Absorption Components in NGC~4151}
\tablewidth{0pt}
\tablehead{
\colhead{Component} & \colhead{Centroid} & \colhead{FWHM} \\
\colhead{} &\colhead{(km s$^{-1}$)} &\colhead{(km s$^{-1}$)} \\
}
\startdata
A     &$-$1588 ($\pm$8) & 36 ($\pm$4) \\
B     &$-$1025 ($\pm$5) & --------$^{a}$ \\
C     &$-$858 ($\pm$8)  & 27 ($\pm$3) \\
D$+$E &$-$491 ($\pm$8)  &435 ($\pm$12) \\
D$'$  &$-$1680 ($\pm$85)&940 ($\pm$60) \\
E$'$  &$-$215 ($\pm$11) &59 ($\pm$8) \\
F     &$-$16 ($\pm$4) & 34 ($\pm$4) \\
F$'$  &$+$34 ($\pm$4) & 15 ($\pm$1) \\

\tablenotetext{a}{sensitive to saturation level, see Danks et al. (2000).}
\enddata
\end{deluxetable}

\clearpage
\begin{deluxetable}{lllllllll}
\tablecolumns{9}
\footnotesize
\tablecaption{Comparison of Component Ionic Columns (x 10$^{13}$ cm$^{-2}$) from 
STIS and GHRS Spectra$^{a}$
\label{tbl-2}}
\tablewidth{0pt}
\tablehead{
\colhead{} &
\colhead{GHRS1} &
\colhead{GHRS2} & 
\colhead{GHRS3} &
\colhead{GHRS4} & 
\colhead{GHRS5} &
\colhead{STIS1} &
\colhead{STIS2}\\
\colhead{} &
\colhead{22.06.92$^{b}$} &
\colhead{04.07.92} &
\colhead{03.01.94} & 
\colhead{28.10.94} &
\colhead{11.03.96} & 
\colhead{25.05.97} &
\colhead{15.07.99}
}
\startdata
Line & & & Comp. A & & & &\\
\hline
Si~II $\lambda$1526 & 1.22 & & & & & $<$0.20 & 0.46$^{c}$\\
                    & ($\pm$0.22) & & & & & & ($\pm$0.08)\\
Si~II$^{*}$ $\lambda$1533 & 1.17 & 0.97 & 1.05 & 1.27 & 0.51 
& $<$0.20 & 0.53$^{d}$\\
                    & ($\pm$0.22) & ($\pm$0.24) & ($\pm$0.19) & ($\pm$0.28) 
& ($\pm$0.16) & & ($\pm$0.08)\\
C~IV $\lambda$1548 & & 2.83 & 3.62 & 3.90 & 3.84 
& 3.13 & 4.11\\
      & & ($\pm$0.31) & ($\pm$0.34) & ($\pm$0.17) & ($\pm$0.22) & ($\pm$0.18) 
& ($\pm$0.31)\\
Mg~II $\lambda$2796 & & & 0.46 & 0.48 & 0.29 
&  & 0.27\\
      & & & ($\pm$0.03) & ($\pm$0.04) & ($\pm$0.03) & & ($\pm$0.04)\\
\hline
Line & & & Comp. C & & & &\\
\hline
Si~II $\lambda$1526 & 0.29 & & & & & 0.32 & 0.46$^{c}$\\
                    & ($\pm$0.13) & & & & & ($\pm$0.12) & ($\pm$0.06)\\
Si~II$^{*}$ $\lambda$1533 & $<$0.20 & $<$0.20 & $<$0.20 & $<$0.20 & $<$0.20
& $<$0.20 & $<$0.20 \\
                    & & & & & & &\\
C~IV $\lambda$1548 & & 2.21 & 2.65 & 2.79 & 2.67 
& 2.50 & 2.95\\
      & & ($\pm$0.37) & ($\pm$0.32) & ($\pm$0.41) & ($\pm$0.25) & ($\pm$0.43) 
& ($\pm$0.40)\\
Mg~II $\lambda$2796 & & & 0.09 & 0.11 & 0.12
&  & 0.10\\
      & & & ($\pm$0.03) & ($\pm$0.02) & ($\pm$0.03) & & ($\pm$0.03)\\
\hline
\tablebreak
Line & & & Comp. E$^{'}$ & & & &\\
\hline
Si~II $\lambda$1526 & $<$0.20 & $<$0.20 & $<$0.20 & 0.34 & $<$0.20 & $<$0.20 
& 1.37\\
                    &  & & & ($\pm$0.11) & & & ($\pm$0.19)\\
Si~II$^{*}$ $\lambda$1533 & $<$0.20 & $<$0.20  & $<$0.20 & 0.63 & $<$0.20
& $<$0.20 & 2.45 \\
                    & & & & ($\pm$0.17) & & & ($\pm$0.31) \\
C~IV $\lambda$1548 & $<$0.30 & $<$0.30 & $<$0.30 & $<$0.30 & $<$0.30
& $<$0.30 & $<$0.30\\
      & & & & & & & \\
Mg~II $\lambda$2803 & & & 0.61 & 0.68 & 0.64
&  & 1.90\\
      & & & ($\pm$0.09) & ($\pm$0.06) & ($\pm$0.06) & & ($\pm$0.13)\\
\tablenotetext{a}{blank entry: not in wavelength coverage. 
Values in parentheses are the measurement errors for the previous
line}
\tablenotetext{b}{Observation date (UT) in Day.Month.Year.}
\tablenotetext{c}{from Si~II $\lambda$1260.4.}
\tablenotetext{d}{from Si~II$^{*}$ $\lambda$1264.7.}
\enddata
\end{deluxetable}

\clearpage
\begin{deluxetable}{lll}
\tablecolumns{3}
\footnotesize
\tablecaption{Predicted and Observed Ionic Columns (x 10$^{13}$ cm$^{-2}$) for Component A
\label{tbl-3}}
\tablewidth{0pt}
\tablehead{
\colhead{Ion (Line)} &
\colhead{Model$^{a}$ Prediction} & 
\colhead{Measured Column}
}
\startdata
H~I (L-$\alpha$) & 8.12E+02 & $>$ 1.0E+01$^{b}$ \\
C~II ($\lambda$1334) & & 3.43 ($\pm$0.53) \\
C~II$^{*}$ ($\lambda$1336) & & 4.19 ($\pm$0.46)  \\
C~II (total)  & 7.42 & 7.62 ($\pm$0.70) \\
C~III & 3.25E+01 & \\
C~IV ($\lambda$1548) & 3.84 & 4.11 ($\pm$0.31)  \\
Mg~II ($\lambda$2796) & 0.21 & 0.27 ($\pm$0.03) \\
Mg~II ($\lambda$2803) & 0.21 & 0.29 ($\pm$0.05) \\
Si~II ($\lambda$1260) &  & 0.46 ($\pm$0.08) \\
Si~II$^{*}$ ($\lambda$1265) & & 0.53 ($\pm$0.15) \\
Si~II (total) & 0.93 & 0.99 ($\pm$0.17) \\
Si~III ($\lambda$1206)& 2.09 & 0.55 ($\pm$0.07)  \\
Si~IV ($\lambda$1393) & 0.80 & 0.98 ($\pm$0.12) \\
Si~IV ($\lambda$1403) & 0.80 & 1.12 ($\pm$0.22)  \\
Fe~II & 0.02  &  \\
\tablenotetext{a}{U $=$ 0.0012, N$_{eff}$ $=$ 1.3E+18 cm$^{-2}$.}
\tablenotetext{b}{Saturated.}
\enddata
\end{deluxetable}

\clearpage
\begin{deluxetable}{lll}
\tablecolumns{3}
\footnotesize
\tablecaption{Predicted and Observed Ionic Columns (x 10$^{13}$ cm$^{-2}$) for Component C
\label{tbl-4}}
\tablewidth{0pt}
\tablehead{
\colhead{Ion (Line)} &
\colhead{Model$^{a}$ Prediction} & 
\colhead{Measured Column}
}
\startdata
H~I (L-$\alpha$)  & 6.40E+03 & \\
C~II ($\lambda$1334) & 5.84 & 4.71 ($\pm$0.36)  \\
C~III & 2.50E+01 & \\
C~IV ($\lambda$1548) & 2.88 & 2.95 ($\pm$0.40) \\
C~IV ($\lambda$1551) & 2.88 & 3.20 ($\pm$0.44) \\
Mg~II ($\lambda$2796) & 0.17 & 0.07 ($\pm$0.01) \\
Mg~II ($\lambda$2803) & 0.17 & 0.10 ($\pm$0.03) \\
Si~II ($\lambda$1260) & 0.76 & 0.30 ($\pm$0.06)  \\
Si~III ($\lambda$1206) & 1.57 & 0.58 ($\pm$0.05) \\
Si~IV ($\lambda$1393) & 0.61 & 0.51 ($\pm$0.07) \\
Si~IV ($\lambda$1403) & 0.61 & 0.70 ($\pm$0.17)  \\
\tablenotetext{a}{U $=$ 0.0012, N$_{eff}$ $=$ 1.0E+18 cm$^{-2}$.}
\enddata
\end{deluxetable}

\clearpage
\begin{deluxetable}{llll}
\tablecolumns{4}
\footnotesize
\tablecaption{Predicted and Observed Ionic Columns (x 10$^{13}$ cm$^{-2}$) for Component D $+$ E
\label{tbl-5}}
\tablewidth{0pt}
\tablehead{
\colhead{Ion (Line$^{a}$)} &
\colhead{Model$^{b}$ Prediction} & 
\colhead{Highstate$^{c}$ Prediction} & 
\colhead{Measured Column}
}
\startdata
H~I (L-$\alpha$)  & 6.4E+06 & 1.6E+04 & \\
C~II & 2.0E+04 & 1.43E+01 &  \\
C~III$^{*}$ $\lambda$1175 & 3.9E+02$^{d}$ & 5.0E+01 & $>$ 3.8E+01 \\
C~IV & 2.5E+04 & 3.1E+04 & $>$ 4.3E+02\\
N~V & 1.5E+03 & 7.9E+03 & $>$ 2.7E+02 \\
O~I & 2.9E+04 & --- &  \\
Mg~II & 1.6E+03 & 6.3 & $>$ 3.4E+01 \\
Al~III & 1.1E+02 & 3.4 &  $>$ 1.2E+02 \\
Si~II ($\lambda$1526) & & & $>$ 1.3E+02  \\
Si~II$^{*}$ ($\lambda$1533) & & & $>$ 8.8E+01 \\
Si~II (total) & 2.1E+03 & 0.21 & $>$ 2.2E+02 \\
Si~III ($\lambda$1206) & 1.7E+03 & 2.7E+01 & $>$ 1.6E+02 \\
Si~IV & 1.9E+03 & 2.1E+02 & $>$ 1.1E+02 \\
P~V & 7.7 & 6.5 & 1.2E+01$^{e}$\\
S~II & 6.3E+02 & --- & 1.0E+03 ($\pm$5.0E+02) \\
Fe~II ($\lambda$1608) & 2.3E+03 & --- & $>$ 1.7E+02$^{f}$ \\
Ni~II & 1.1E+02 & --- & 8.5E+01 ($\pm$2.2E+01)\\
\tablenotetext{a}{when no wavelength is listed, the measured
column is derived from a simultaneous fit to two or more lines
(see text).}
\tablenotetext{b}{U $=$ 0.015, N$_{eff}$ $=$ 2.75E+21 cm$^{-2}$.}
\tablenotetext{c}{U $=$ 0.060, N$_{eff}$ $=$ 2.75E+21 cm$^{-2}$.}
\tablenotetext{d}{assuming N$_{metastable}$/N$_{ground}$ $=$ 0.01 (see 
text).}
\tablenotetext{e}{value from {\it ORFEUS-SPAS II} observation (Espey et al.
1998).}
\tablenotetext{f}{determined from resonance lines.}
\enddata
\end{deluxetable}

\clearpage
\begin{deluxetable}{llll}
\tablecolumns{4}
\footnotesize
\tablecaption{Predicted and Observed Ionic Columns (x 10$^{13}$ cm$^{-2}$) for Component D$^{'}$
\label{tbl-6}}
\tablewidth{0pt}
\tablehead{
\colhead{Ion (Line$^{a}$)} &
\colhead{Model$^{b}$ Prediction} & 
\colhead{Highstate$^{c}$ Prediction} & 
\colhead{Measured Column}
}
\startdata
H~I (L-$\alpha$)  & 3.34E+04 & 3.33 \\
C~II & 8.54E+01 & --- &   \\
C~III & 3.83E+02 & --- & \\
C~IV & 2.58E+02 & 0.19 & 1.20E+02 ($\pm$2.40E+01)\\
N~V & 1.70E+02 & 0.97 & 2.20E+02 ($\pm$4.40E+01)\\
Mg~II & 1.30 & --- &  \\
Si~II & 3.26 & --- &  \\
Si~III & 2.15E+01 & --- &  \\
Si~IV & 2.82E+01 & --- & 4.50E+01 ($\pm$9.00)\\
\tablenotetext{a}{when no wavelength is listed, the measured
column is derived from a simultaneous fit to two or more lines
(see text).}
\tablenotetext{b}{screened by Component D$+$E (see text), U $=$ 0.012,
N$_{eff}$ $=$ 1.0E+20 cm$^{-2}$.}
\tablenotetext{c}{screened by Highstate Component D$+$E (see text), U $=$ 4.15,
N$_{eff}$ $=$ 1.0E+20 cm$^{-2}$.}
\enddata
\end{deluxetable}

\clearpage
\begin{deluxetable}{lll}
\tablecolumns{3}
\footnotesize
\tablecaption{Predicted and Observed Ionic Columns (x 10$^{13}$ cm$^{-2}$) for Component E$^{'}$
\label{tbl-7}}
\tablewidth{0pt}
\tablehead{
\colhead{Ion (Line)} &
\colhead{Model Prediction$^{a}$} & 
\colhead{Measured Column}
}
\startdata
H~I (L-$\alpha$)  & 8.83E+03  & \\
C~II ($\lambda$1334) & & 9.73 ($\pm$3.18) \\
C~II$^{*}$ ($\lambda$1335) & & 1.23E+01 ($\pm$3.11) \\
C~II (total) & 3.56E+01 & 2.20E+01 ($\pm$4.45)  \\
C~III & 1.46E+01 & \\
O~I ($\lambda$1302) & & 5.47 ($\pm$2.54) \\
O~I$^{*}$ ($\lambda$1304) & & 3.48 ($\pm$1.54) \\
O~I (total) & 6.26 & 8.95 ($\pm$2.97) \\
Mg~II ($\lambda$2803) & 2.03 & 1.90 ($\pm$0.13) \\
Si~II ($\lambda$1526) & & 1.37 ($\pm$0.19)  \\
Si~II$^{*}$ ($\lambda$1533) & & 2.45 ($\pm$0.31) \\
Si~II (total) & 3.75 & 3.82 ($\pm$0.36) \\
Si~III ($\lambda$1206) & 0.86 & \\
Fe~II ($\lambda$2383) & 2.31 & 0.50 ($\pm$0.11) \\
Fe~II ($\lambda$2600) & 2.31 & 0.31 ($\pm$0.13)   \\
\tablenotetext{a}{U $=$ 0.0001, N$_{eff}$ $=$ 1.5E+18 cm$^{-2}$.}
\enddata
\end{deluxetable}

\clearpage
\vskip3.0in
\begin{figure}
\plotone{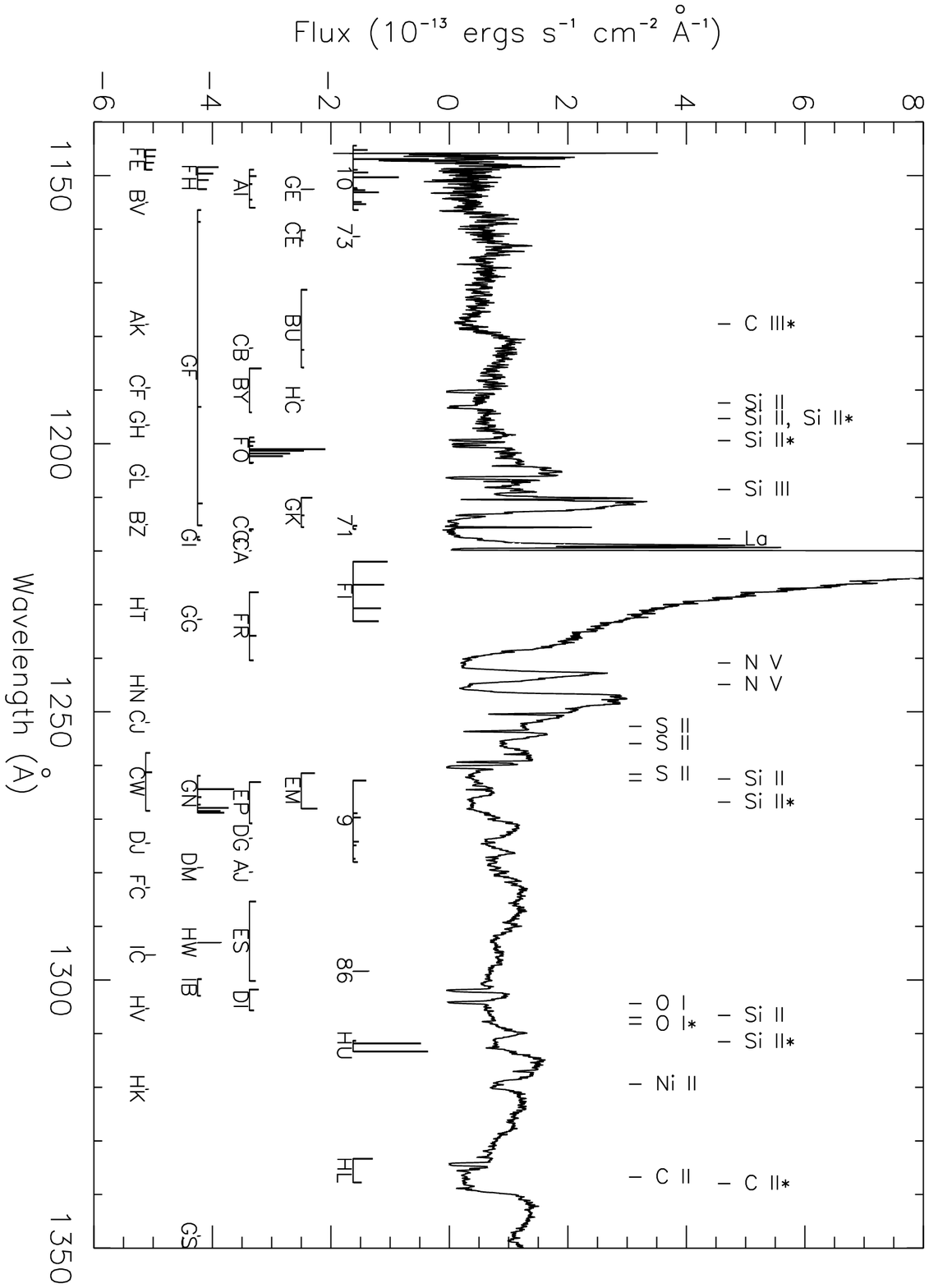}
\\Fig.~1a.
\end{figure}

\clearpage
\vskip3.0in
\begin{figure}
\plotone{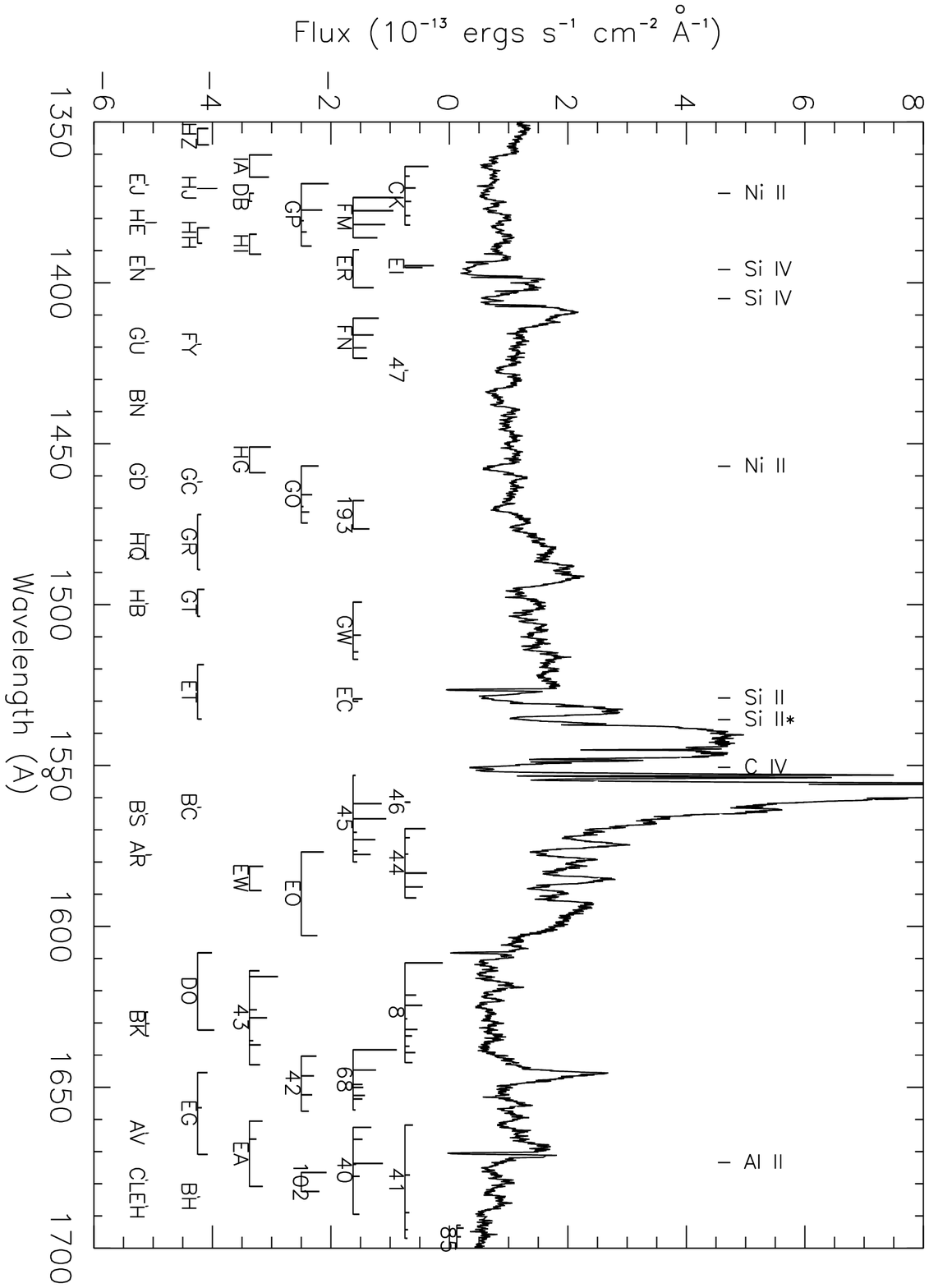}
\\Fig.~1b.
\end{figure}

\clearpage
\vskip3.0in
\begin{figure}
\plotone{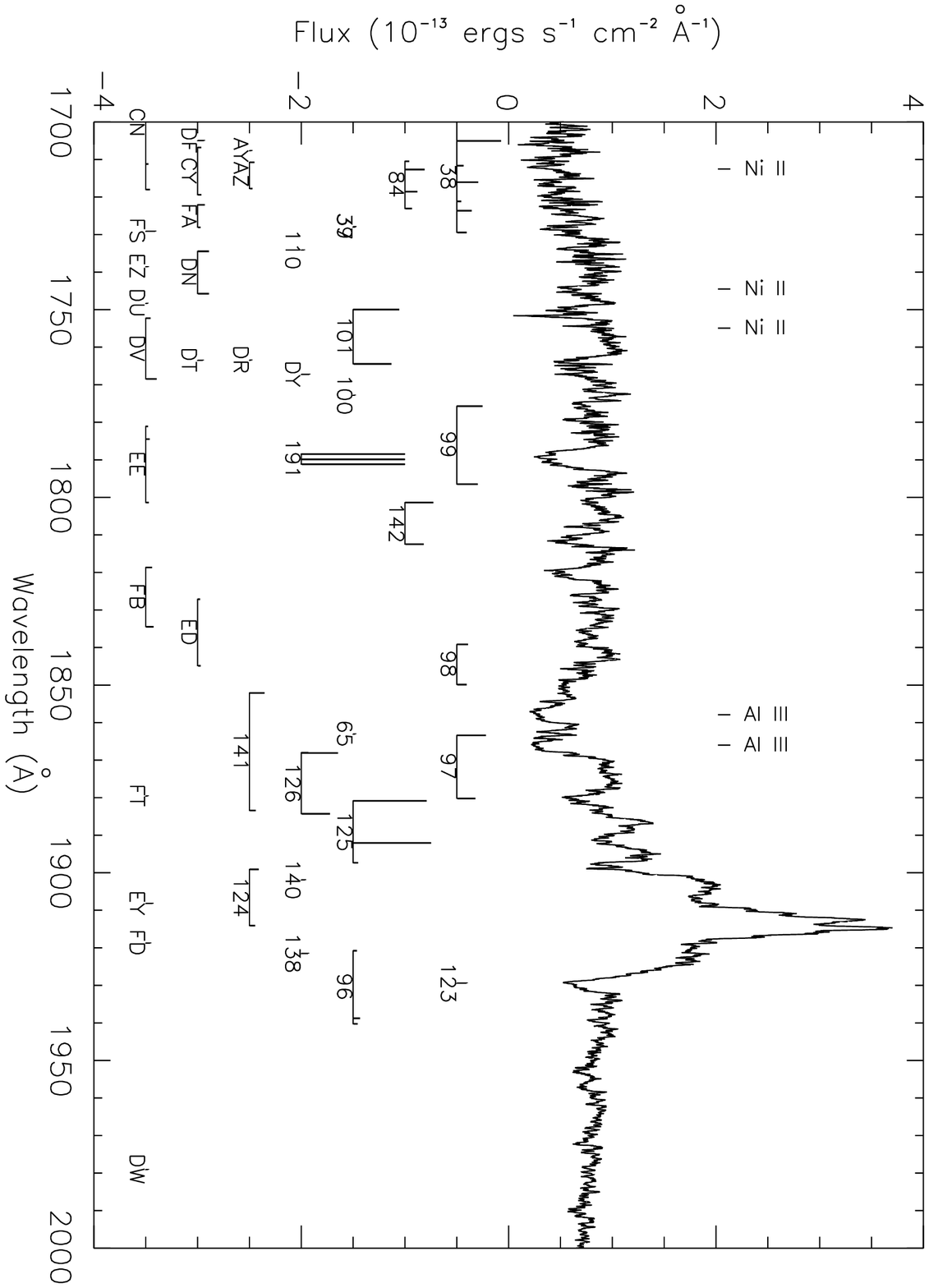}
\\Fig.~1c.
\end{figure}

\clearpage
\vskip3.0in
\begin{figure}
\plotone{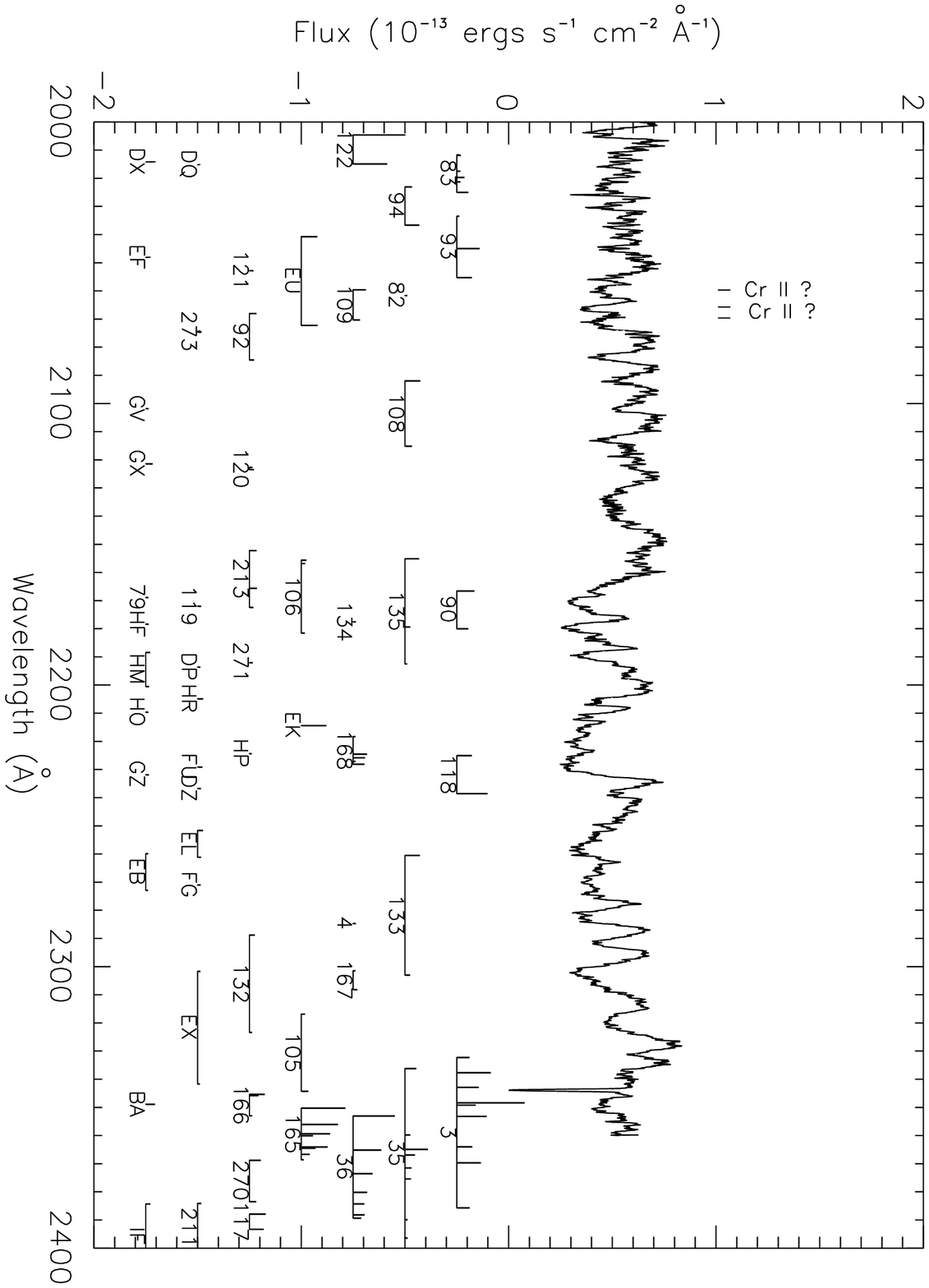}
\\Fig.~1d.
\end{figure}

\clearpage
\vskip3.0in
\begin{figure}
\plotone{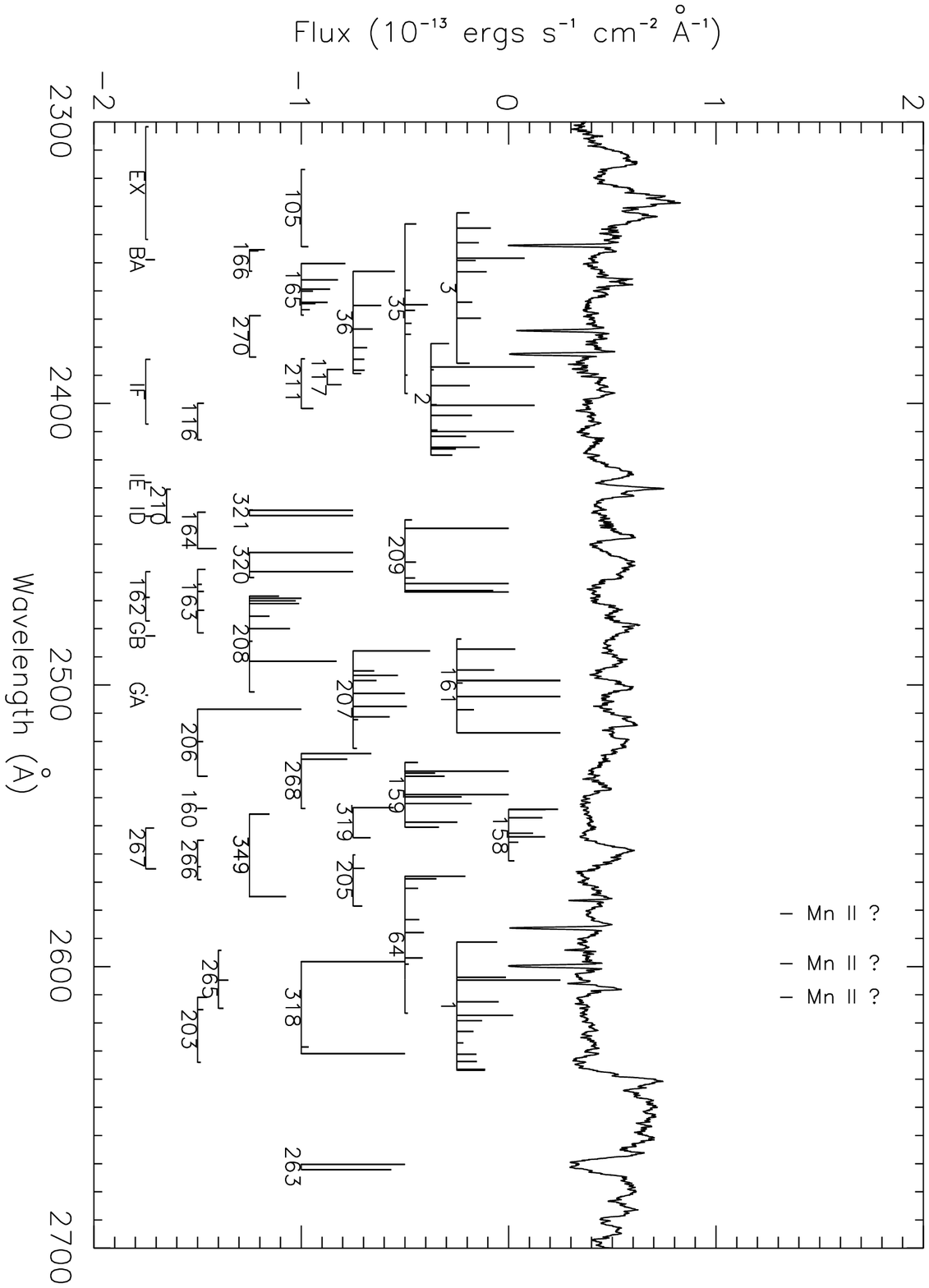}
\\Fig.~1e.
\end{figure}

\clearpage
\vskip3.0in
\begin{figure}
\plotone{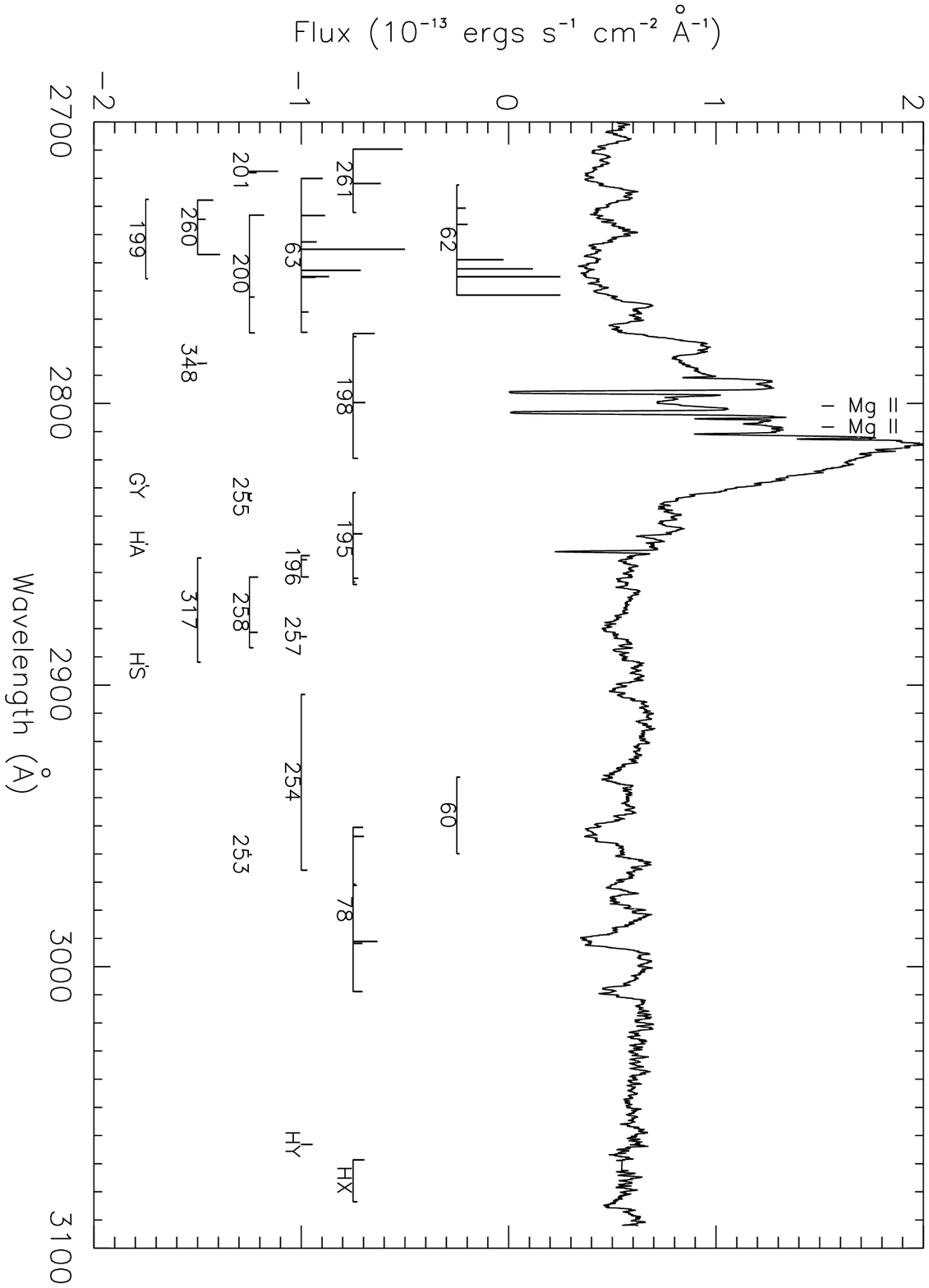}
\\Fig.~1f.
\end{figure}

\clearpage
\vskip3.0in
\begin{figure}
\plotone{fig2.ps}
\\Fig.~2.
\end{figure}

\clearpage
\vskip3.0in
\begin{figure}
\plotone{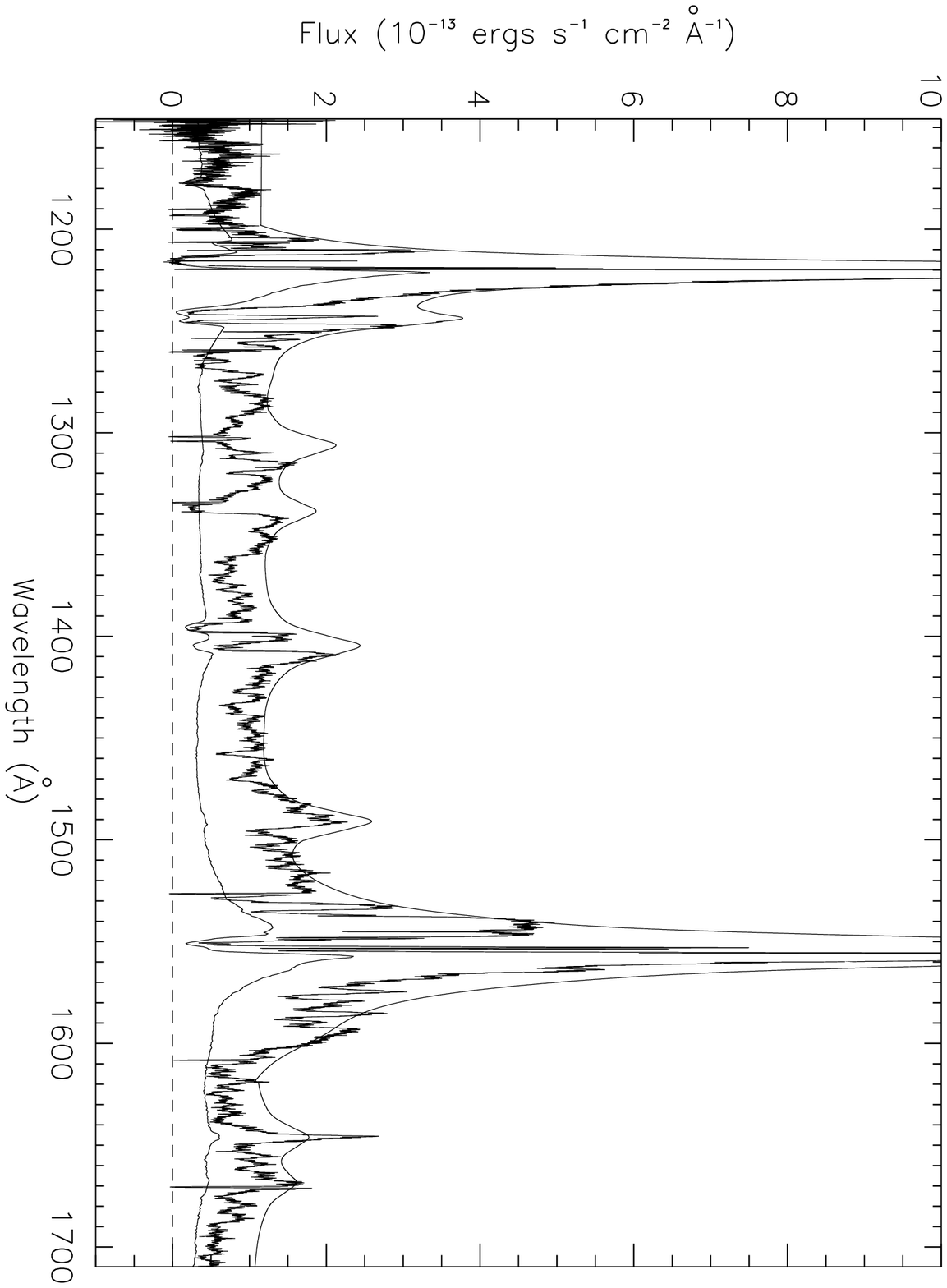}
\\Fig.~3a.
\end{figure}

\clearpage
\vskip3.0in
\begin{figure}
\plotone{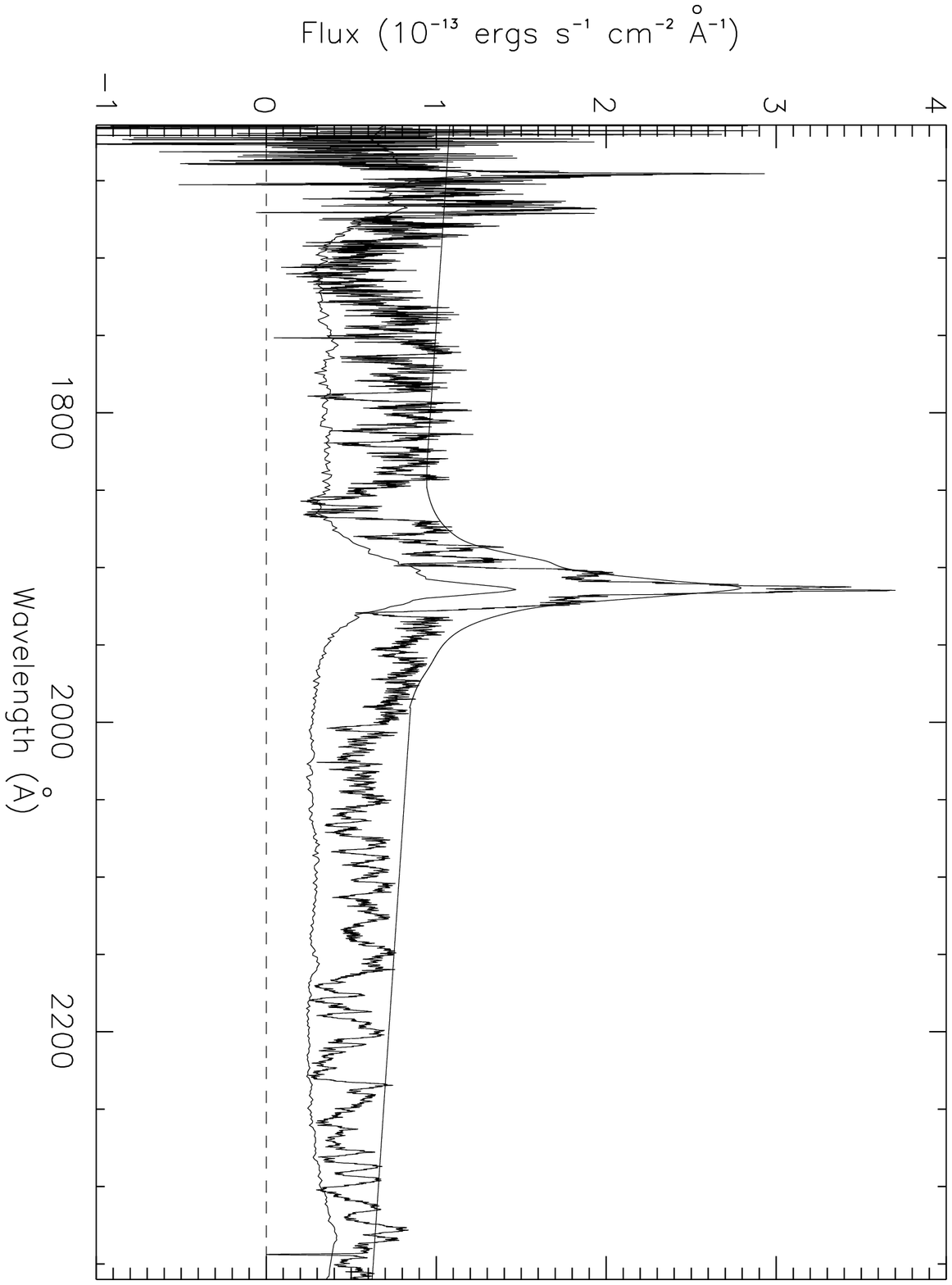}
\\Fig.~3b.
\end{figure}

\clearpage
\vskip3.0in
\begin{figure}
\plotone{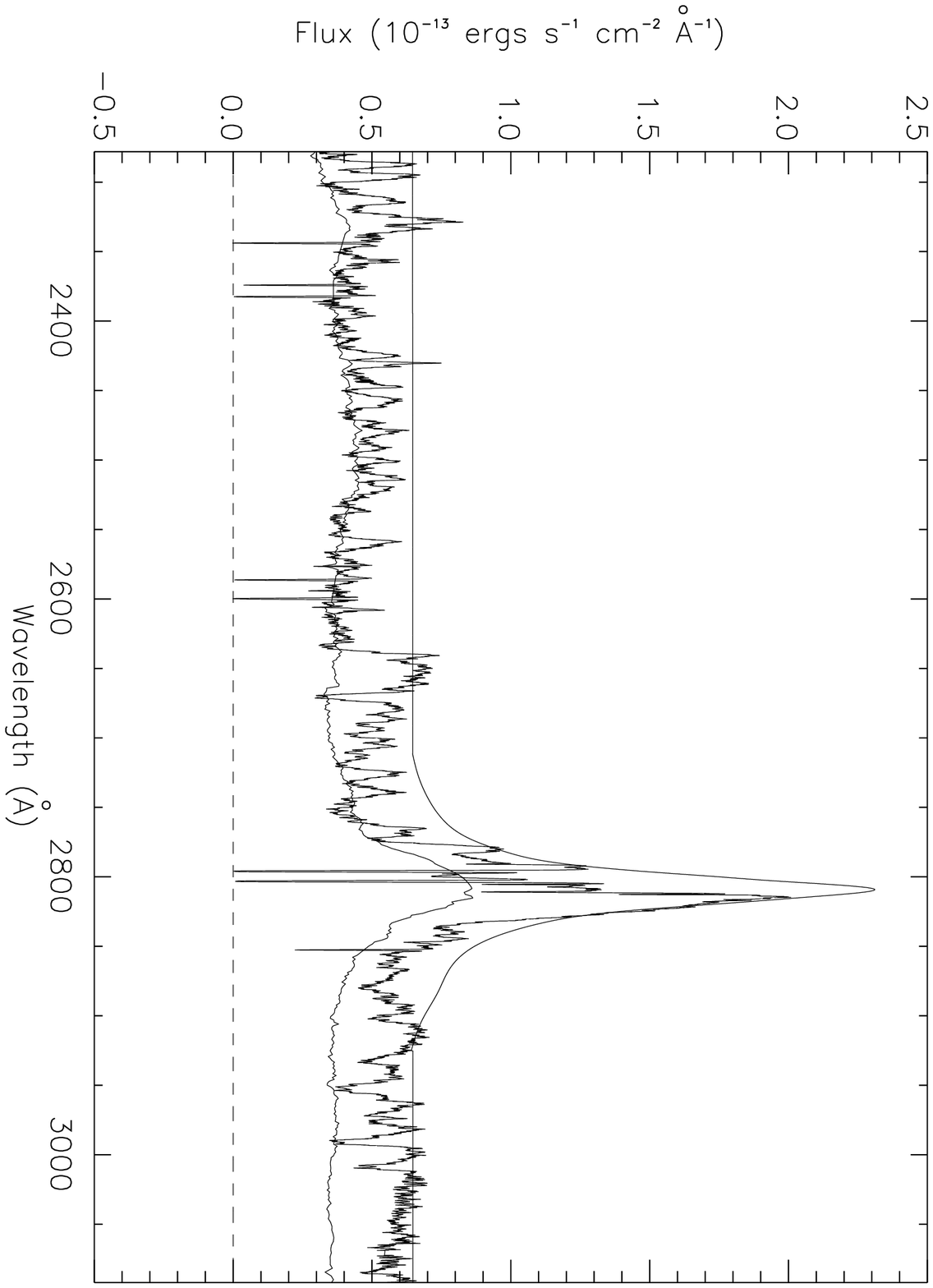}
\\Fig.~3c.
\end{figure}

\clearpage
\vskip3.0in
\begin{figure}
\plotone{fig4.ps}
\\Fig.~4.
\end{figure}

\clearpage
\vskip3.0in
\begin{figure}
\plotone{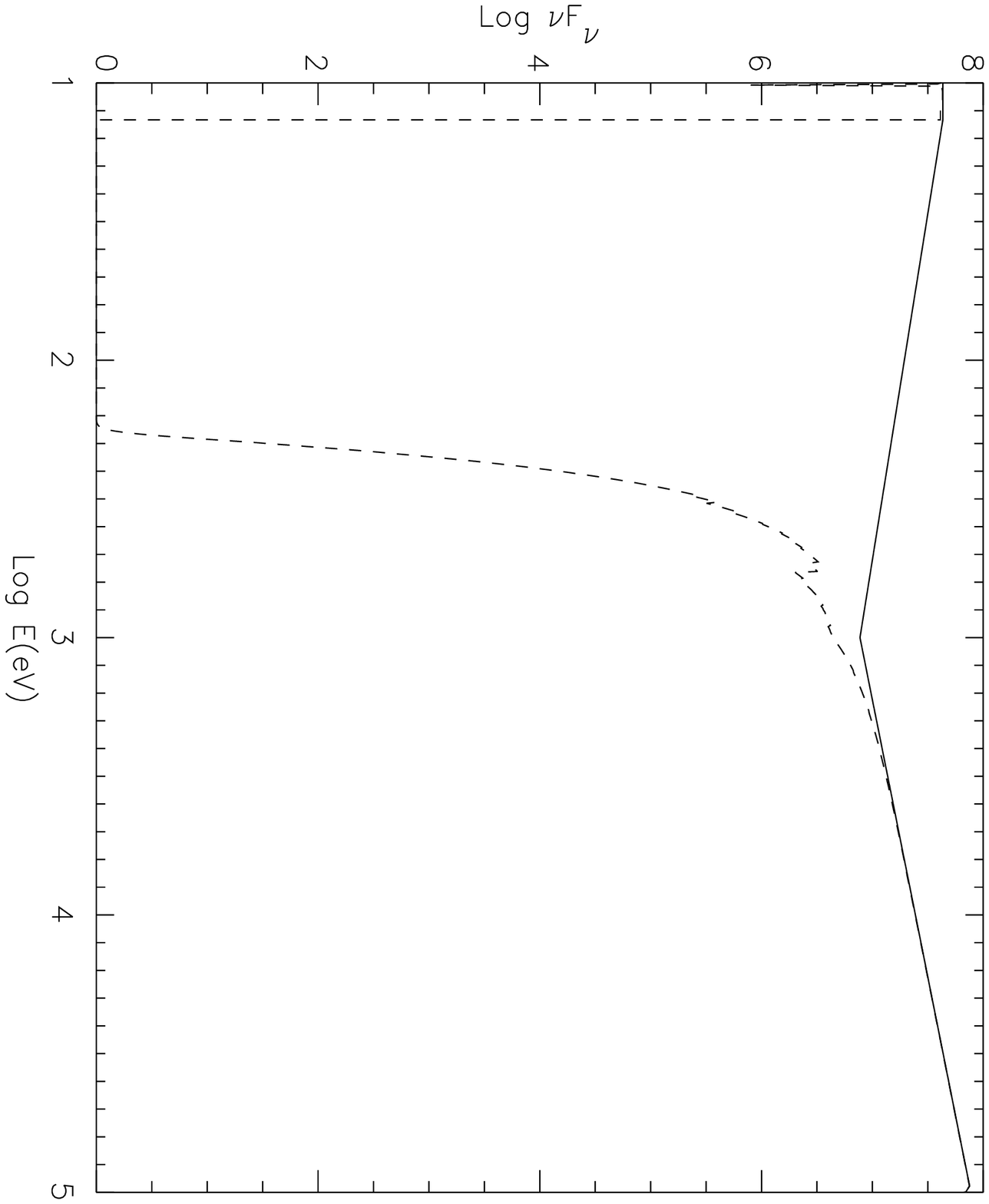}
\\Fig.~5a.
\end{figure}

\clearpage
\vskip3.0in
\begin{figure}
\plotone{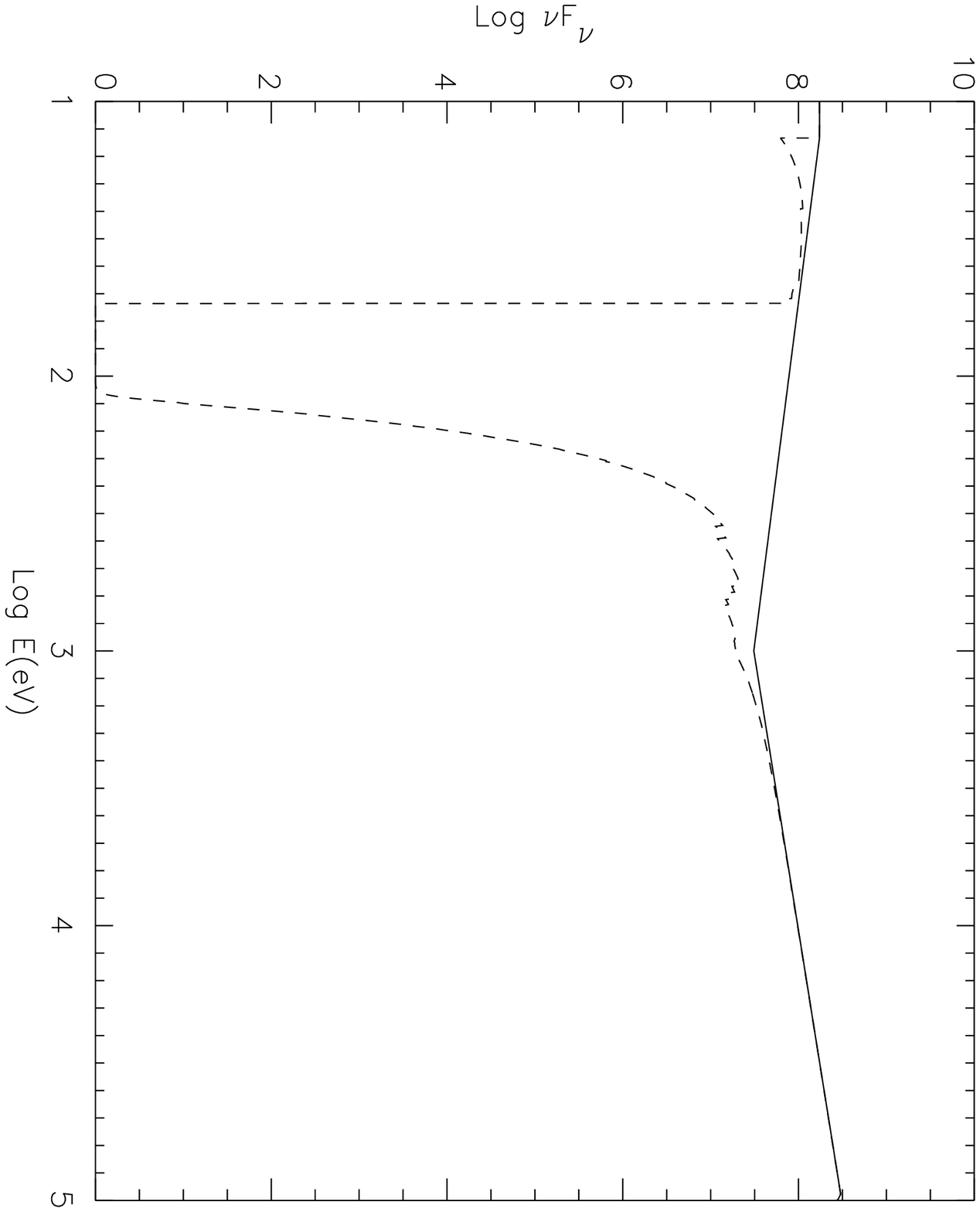}
\\Fig.~5b.
\end{figure}

\clearpage
\vskip3.0in
\begin{figure}
\plotone{fig6.ps}
\\Fig.~6.
\end{figure}

\end{document}